\documentclass{svproc}
\usepackage{url}

\usepackage[printonlyused, nolist]{acronym}
\usepackage{amsmath}
\usepackage{amsfonts}
\usepackage{graphicx}
\usepackage{mathtools}
\usepackage{subcaption}
\usepackage{hyperref}

\usepackage{todonotes}

\begin{acronym}
  \acro{qaoa}[QAOA]{Quantum Approximate Optimization Algorithm}
  \acro{gmvp}[GMVP]{Generalized Mean-Variance Problems}
  \acro{nisq}[NISQ]{Noisy Intermediate-Scale Quantum}
  \acro{vqas}[VQAs]{Variational Quantum Algorithms}
  \acro{cobyla}[COBYLA]{Constrained Optimization by Linear Approximation}
\end{acronym}

\begin{document}
\mainmatter              
\title{Classical Optimization Strategies for Variational Quantum Algorithms: A Systematic Study of Noise Effects and Parameter Efficiency}
\titlerunning{VQA Optimization: Landscape, Noise, and Parameter Efficiency}  
%
\author{
Tom\'{a}\v{s} Bezd\v{e}k\inst{1} \and
Haomu Yuan\inst{2} \and
Vojt\v{e}ch Nov\'{a}k\inst{3,4} \and
Silvie Ill\'{e}sov\'{a}\inst{5} \and
Martin Beseda\inst{6}
}
\authorrunning{Tomáš Bezděk et al.} 
%
\tocauthor{Ivar Ekeland, Roger Temam, Jeffrey Dean, David Grove,
Craig Chambers, Kim B. Bruce, and Elisa Bertino}
\institute{Department of Mathematics, TUM School of Computation, Information and Technology, Technical University of Munich, Garching bei München, Germany\\
\email{tomas.bezdek@tum.de}
\and
Cavendish Laboratory, Department of Physics, University of Cambridge, Cambridge CB3 0US, UK\and
Department of Computer Science, Faculty of Electrical Engineering and Computer Science, VSB - Technical University of Ostrava, Ostrava, Czech Republic\and
IT4Innovations National Supercomputing Center,\\
VSB - Technical University of Ostrava, 708 00 Ostrava, Czech Republic \and
Gran Sasso Science Institute, L’Aquila, Italy \and
Dipartimento di Ingegneria e Scienze dell’Informazione e Matematica,\\ 
Università dell’Aquila, Via Vetoio, I-67010 Coppito, L’Aquila, Italy }

\maketitle              

\begin{abstract}
This study systematically benchmarks classical optimization strategies for the Quantum Approximate Optimization Algorithm when applied to Generalized Mean-Variance Problems under near-term Noisy Intermediate-Scale Quantum conditions. We evaluate Dual Annealing, Constrained Optimization by Linear Approximation, and the Powell Method across noiseless, sampling noise, and two thermal noise models. Our Cost Function Landscape Analysis revealed that the Quantum Approximate Optimization Algorithm angle parameters $\gamma$ were largely inactive in the noiseless regime. This insight motivated a parameter-filtered optimization approach, in which we focused the search space exclusively on the active $\beta$ parameters. This filtering substantially improved parameter efficiency for fast optimizers like Constrained Optimization by Linear Approximation (reducing evaluations from 21 to 12 in the noiseless case) and enhanced robustness, demonstrating that leveraging structural insights is an effective architecture-aware noise mitigation strategy for Variational Quantum Algorithms.
\end{abstract}

\section{Introduction}
\label{sec:intro}
The field of quantum computing is rapidly maturing~\cite{bauer2025efficient,gupta2022quantum,evolvingcircuits,illesova2025qmetric,illesova2025importance,novak2025predicting,rajamani2025equiensembledescriptionsystematicallyoutperforms,zelinka2023isoma,zhang2025qracle}.
On current \ac{nisq} architectures~\cite{bilek2025experimental,lewandowska2025benchmarking}, \ac{vqas}~\cite{beseda2024state,PhysRevA.111.022437,illesova2025transformation,trovato2025preliminary} are a leading paradigm; yet, their performance is critically dependent on the classical optimization of ansatz parameters~\cite{Cerezo2021}.
This process is severely challenged by hardware noise, which degrades the optimization landscape and can obscure gradients~\cite{vha,saoovqe_benchmark,novak2025optimization}.
This noise creates a fundamental ambiguity: should one employ sample-efficient but noise-vulnerable gradient-based methods, or gradient-free methods (e.g., \ac{cobyla}~\cite{cobyla}, Dual Annealing~\cite{dual_annealing}) that offer robustness but are often deemed less efficient?

This work addresses the optimal choice of classical optimizers for \ac{qaoa} by presenting two primary contributions.
Firstly, a systematic benchmark of \ac{qaoa} on \ac{gmvp} across four distinct noise models (including two realistic thermal noise profiles) is conducted, incorporating a Cost Function Landscape Analysis to visually assess how noise affects landscape ruggedness and to identify parameter activity.
Secondly, a comparison of classical optimizers benchmarking their standard performance against a parameter-filtered scenario where optimization is restricted to only the active parameters, thereby testing a novel approach for improving efficiency and robustness.

\section{Methods}
\subsection{\ac{qaoa}}
To investigate the general applicability of the \ac{qaoa} for practical problems, we implement the hard-constrained \ac{qaoa} proposed in~\cite{yuan2024quantifyingadvantagesapplyingquantum,yuan2025iterativequantumoptimisationwarmstarted}.
Consider a constrained discrete optimization problem:
\begin{equation}\label{qdp}
    \hspace*{-1.1cm}
    \begin{aligned}
        \underset{\boldsymbol{x}\in\{0,1\}^n}{\operatorname*{argmin}} \quad &
        f(\boldsymbol{x}) = \boldsymbol{x}^{T} \boldsymbol{\Sigma} \boldsymbol{x}
    \end{aligned}
\end{equation}
\vspace{-2.2em}
\begin{subequations}
    \begin{align}
        \quad \text{s.t.} \quad & c(\boldsymbol{x}) = \boldsymbol{x}^{T} \mathbf{1}_n - 1 = 0
    \end{align}
\end{subequations}
where $\boldsymbol{x}$ is an $n$-dimensional binary vector, $\mathbf{1}_n$ is an $n$-dimensional vector of ones, and $\sigma_{ij}\in\boldsymbol{\Sigma}$ denotes the correlation coefficient between variables $i$ and $j$.

The hard-constrained \ac{qaoa} circuit employs a parameterized, alternating structure given by
\begin{equation}\label{altCirc}
    \begin{aligned}
        \hat{U}(c,\beta_p)\, \hat{U}(f,\gamma_p) \cdots \hat{U}(c,\beta_1)\, \hat{U}(f,\gamma_1)\, \hat{U}_I |0\rangle,
    \end{aligned}
\end{equation}
where $\hat{U}_I$ denotes the initial state preparation unitary.
The cost operator $\hat{U}(f,\gamma)$ encodes the objective function as
\[
    \hat{U}(f,\gamma)|x\rangle = e^{-i\gamma \hat{H}_f}|x\rangle = e^{-i\gamma f(x)}|x\rangle,
\]
and the hard-mixing operator $\hat{U}(c,\beta)$ enforces transitions only within the feasible subspace defined by the constraint function. This operator is designed to efficiently utilize qubit excitations and is given by
\begin{equation}\label{eq:QE23}
    \begin{aligned}
        \hat{U}(c,\beta) = \prod_{tt'} \tilde{S}_{tt'}(\beta)\, \tilde{P}^{(1)}_{tt'}(\beta)\, \tilde{P}^{(2)}_{tt'}(\beta)\, \tilde{S}_{tt'}(\beta),
    \end{aligned}
\end{equation}
where
\begin{equation}\label{eq:QE2}
    \begin{aligned}
         & \tilde{S}_{tt'}(\beta) = \bigotimes_{k=1}^l e^{-i\beta \hat{S}^k_{tt'}},
        \quad
        \tilde{P}^{(j)}_{tt'}(\beta) = \bigotimes_{k \in \mathcal{K}_j} e^{-i\beta \hat{P}^{k}_{ttt'}},        \\[0.5em]
         & \hat{S}_{tt'}^k = -\tfrac{1}{2}(\hat{X}^{k}_{t}\hat{Y}^{k}_{t'} - \hat{Y}^{k}_{t}\hat{X}^{k}_{t'}), \\[0.5em]
         & \hat{P}_{ttt'}^{k} = -\tfrac{1}{4}(\hat{X}^{k+1}_{t}\hat{X}^{k}_{t}\hat{Y}^{k}_{t'}
        + \hat{X}^{k+1}_{t}\hat{Y}^{k}_{t}\hat{X}^{k}_{t'}
        - \hat{Y}^{k+1}_{t}\hat{X}^{k}_{t}\hat{X}^{k}_{t'}
        + \hat{Y}^{k+1}_{t}\hat{Y}^{k}_{t}\hat{Y}^{k}_{t'}).
    \end{aligned}
\end{equation}
Here, $\hat{X}$ and $\hat{Y}$ denote the Pauli-X and Pauli-Y operators, respectively.
Indices $t$ and $t'$ correspond to qubit blocks representing binary variables, while $k$ indexes the qubits within each block.
The index sets are defined as
\[
    \mathcal{K}_1 = \{2c_1 \mid c_1 \in [1, \lfloor l/2 \rfloor] \cap \mathbb{Z}\}, \quad
    \mathcal{K}_2 = \{2c_2 - 1 \mid c_2 \in [1, \lceil l/2 \rceil] \cap \mathbb{Z}\},
\]
with $\mathcal{K}_1$ and $\mathcal{K}_2$ reduced modulo $l$, and the difference $t-t'$ taken modulo $n$.

Finally, $p$ denotes the number of \ac{qaoa} layers, and each layer consists of one cost and one mixing operator parameterized by angles $\gamma$ and $\beta$, which
represent the sets of variational parameters.
The parameter ranges are typically restricted to
$\gamma \in [0, 2\pi]$ and $\beta \in [0, \pi]$, respectively.

In the experiments, we used the hard-constrained \ac{qaoa} with $p = 2$ layers to solve a \ac{gmvp} instance with $n = 4$ assets, each encoded with three qubits, for a total of $12$ qubits in the circuit.

\subsection{Optimizers}
\label{sec:optimizers}
We systematically benchmarked three derivative-free classical optimizers selected for their noise robustness and performance profile: Dual Annealing~\cite{dual_annealing} (a global metaheuristic), \ac{cobyla}~\cite{cobyla} (a fast local direct search), and the Powell Method~\cite{10.1007/BFb0067703} (a local trust-region method).

\section{Results}
\subsection{Noise Profiles}
\label{sec:noise_profiles}
We executed our experiments across four distinct noise setups, with 1024 shots used for all noisy profiles. The baseline was the noiseless setup (ideal state vector simulation). This was compared against three noisy regimes: Sampling Noise (statistical noise only), and two full physical noise models: Thermal Noise-A ($T_1=380\mu s,T_2=400\mu s$) and Thermal Noise-B ($T_1=80\mu s,T_2=100\mu s$)~\cite{Georgopoulos_2021}. All gates used execution times of $50ns$ (one-qubit) and $150ns$ (two-qubit), allowing us to distinguish between statistical noise and physical decoherence effects.

\subsection{Landscapes}
\label{sec:landscape}
We performed a Cost Function Landscape Analysis to gain insight into the optimization characteristics. Contour plots were generated for all two-dimensional subspaces defined by pairs of parameters from $\{ \beta_1 , \beta_2, \gamma_1, \gamma_2 \}$, with the remaining two parameters fixed to their known global minimum values ($\theta^* = [0.0, 0.0, 0.14286, 0.85714]$).

\begin{figure}[htb]
    \centering
    \begin{subfigure}[b]{0.28\textwidth}
        \centering
        \includegraphics[width=\textwidth]{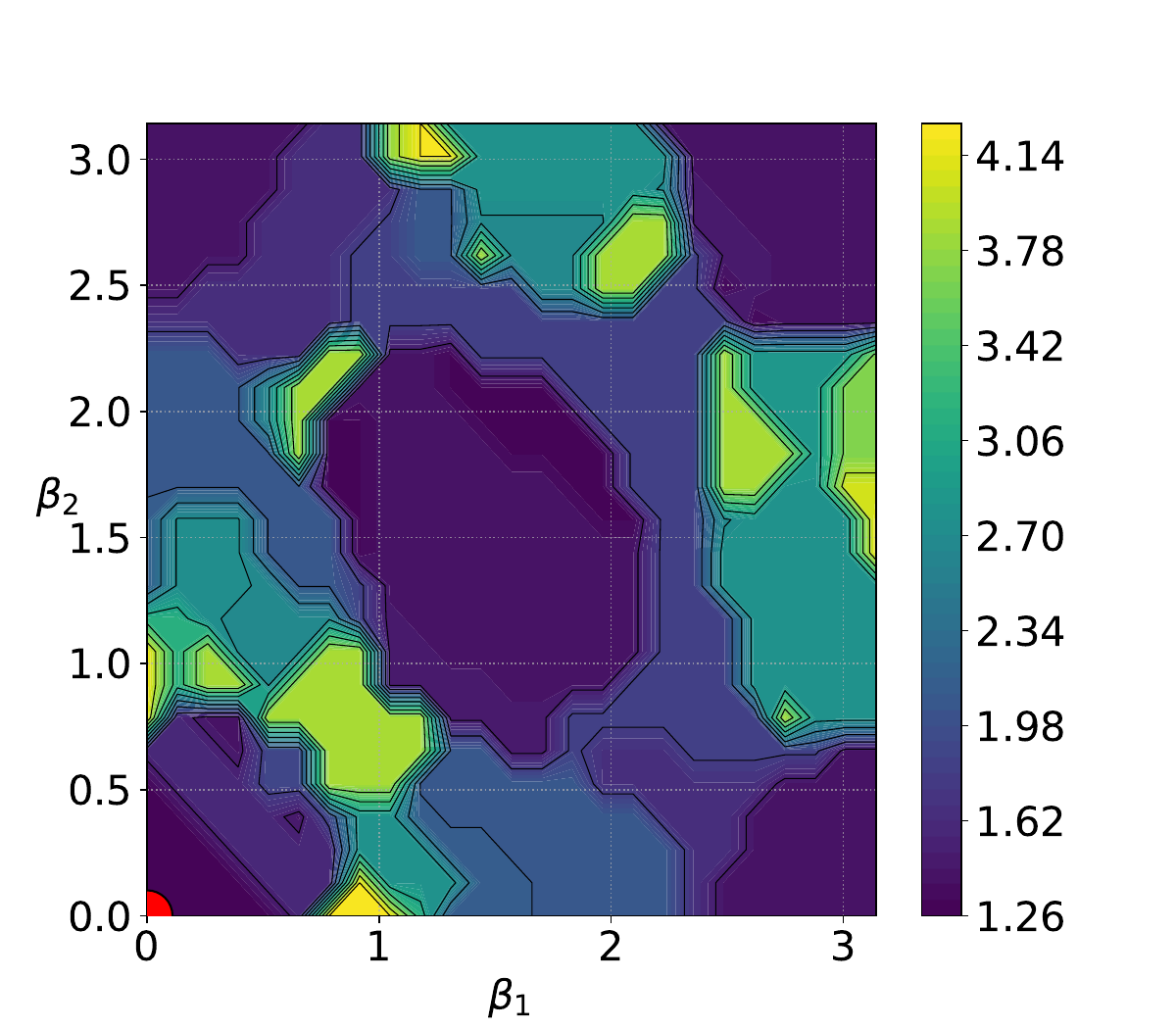}
        \caption{Landscape for $\beta_1$, $\beta_2$}
        \label{fig:sv-a}
    \end{subfigure}
    \hfill 
    \begin{subfigure}[b]{0.28\textwidth}
        \centering
        \includegraphics[width=\textwidth]{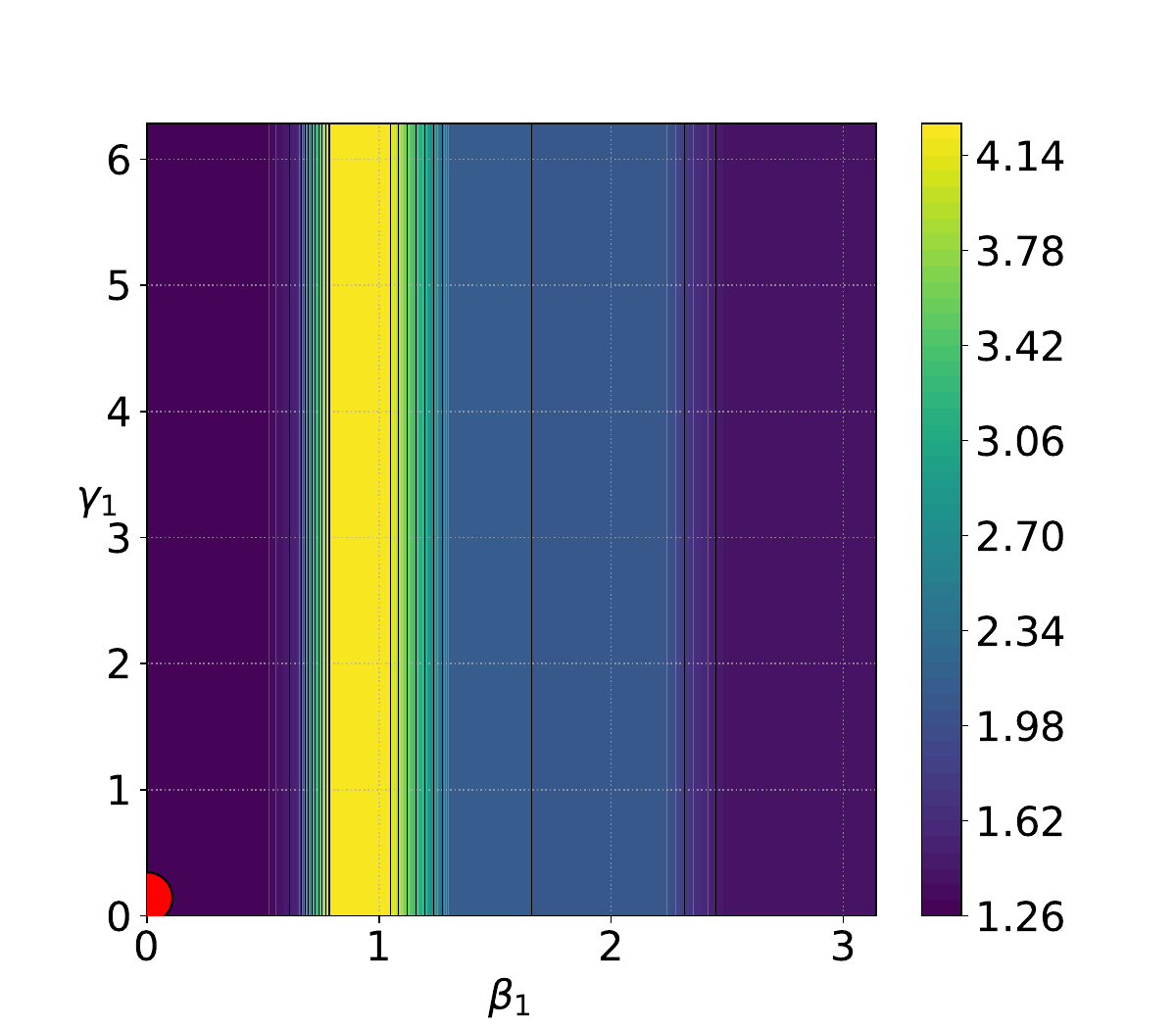}
        \caption{Landscape for $\beta_1$, $\gamma_1$}
        \label{fig:sv-b}
    \end{subfigure}
    \hfill
    \begin{subfigure}[b]{0.28\textwidth}
        \centering
        \includegraphics[width=\textwidth]{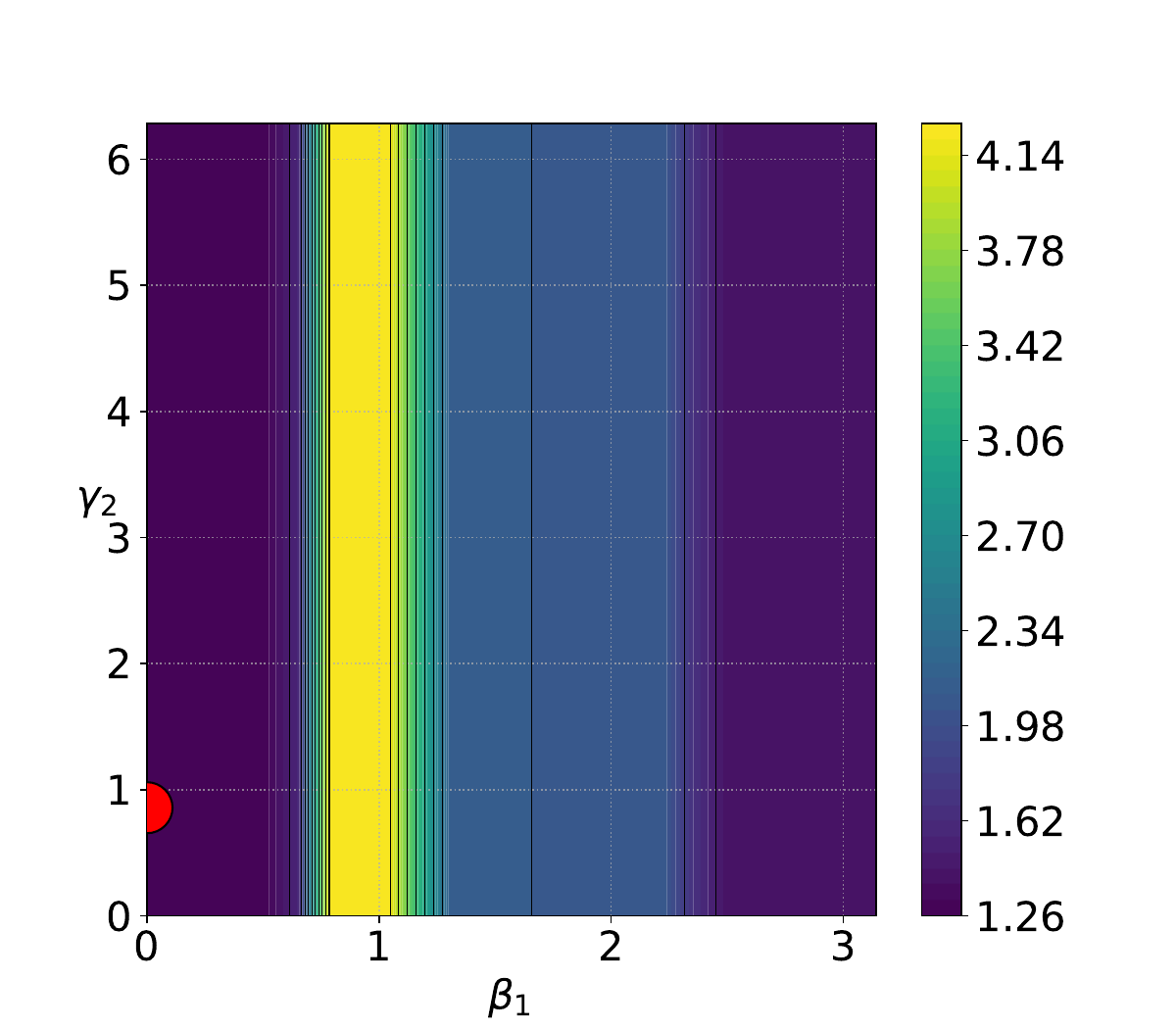}
        \caption{Landscape for $\beta_1$, $\gamma_2$}
        \label{fig:sv-c}
    \end{subfigure}

    \begin{subfigure}[b]{0.28\textwidth}
        \centering
        \includegraphics[width=\textwidth]{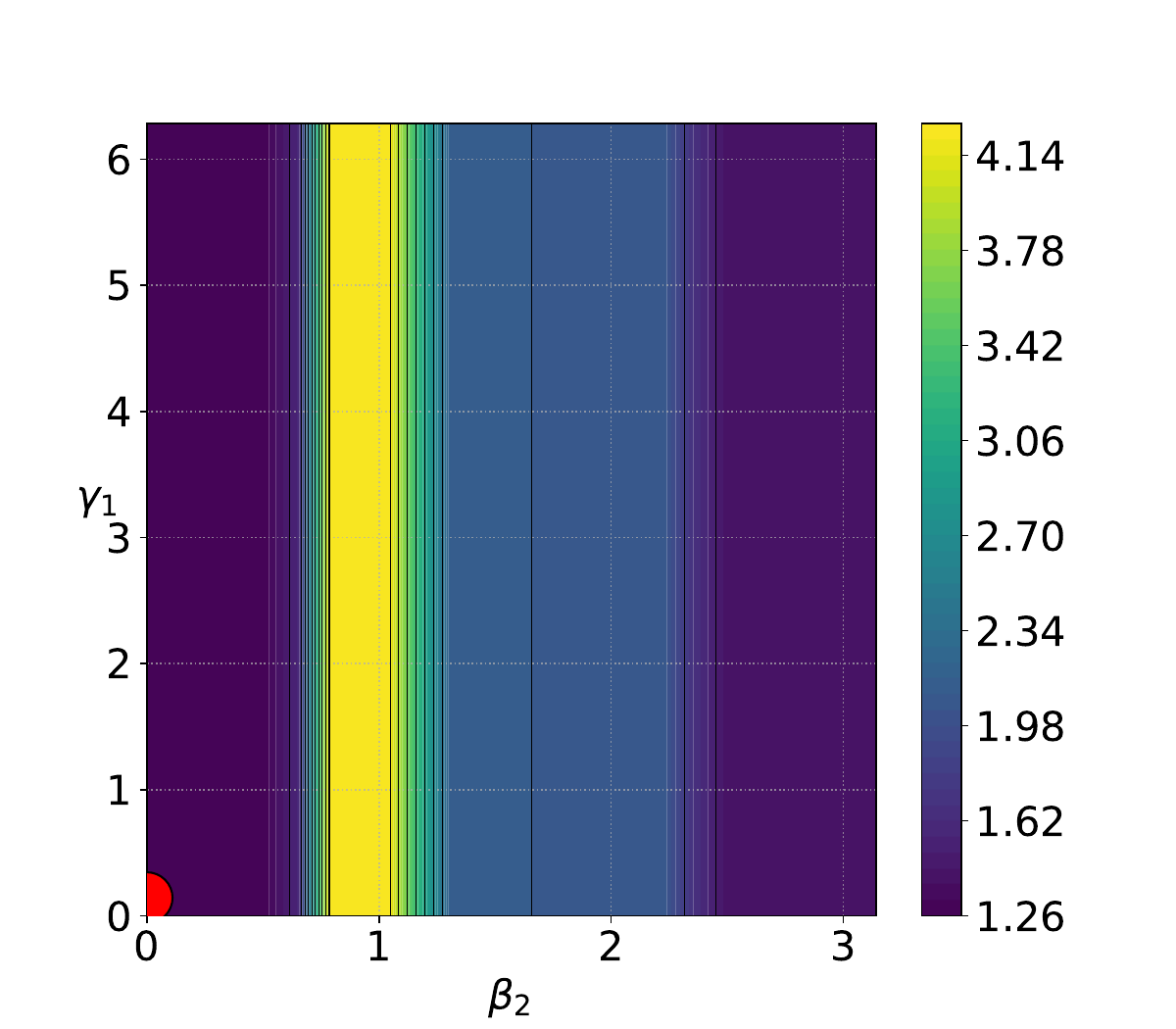}
        \caption{Landscape for $\beta_2$, $\gamma_1$}
        \label{fig:sv-d}
    \end{subfigure}
    \hfill 
    \begin{subfigure}[b]{0.28\textwidth}
        \centering
        \includegraphics[width=\textwidth]{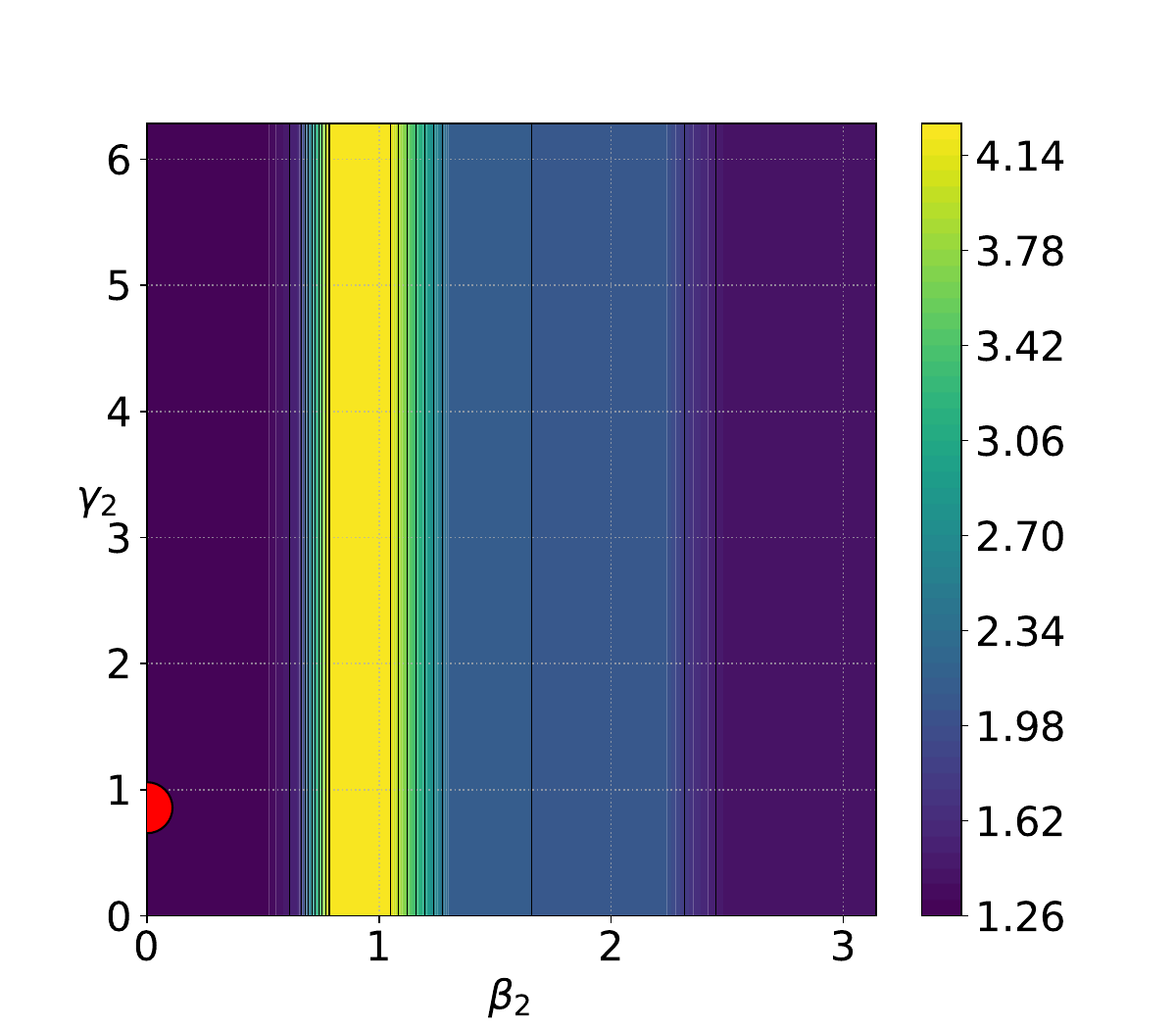}
        \caption{Landscape for $\beta_2$, $\gamma_2$}
        \label{fig:sv-e}
    \end{subfigure}
    \hfill
    \begin{subfigure}[b]{0.28\textwidth}
        \centering
        \includegraphics[width=\textwidth]{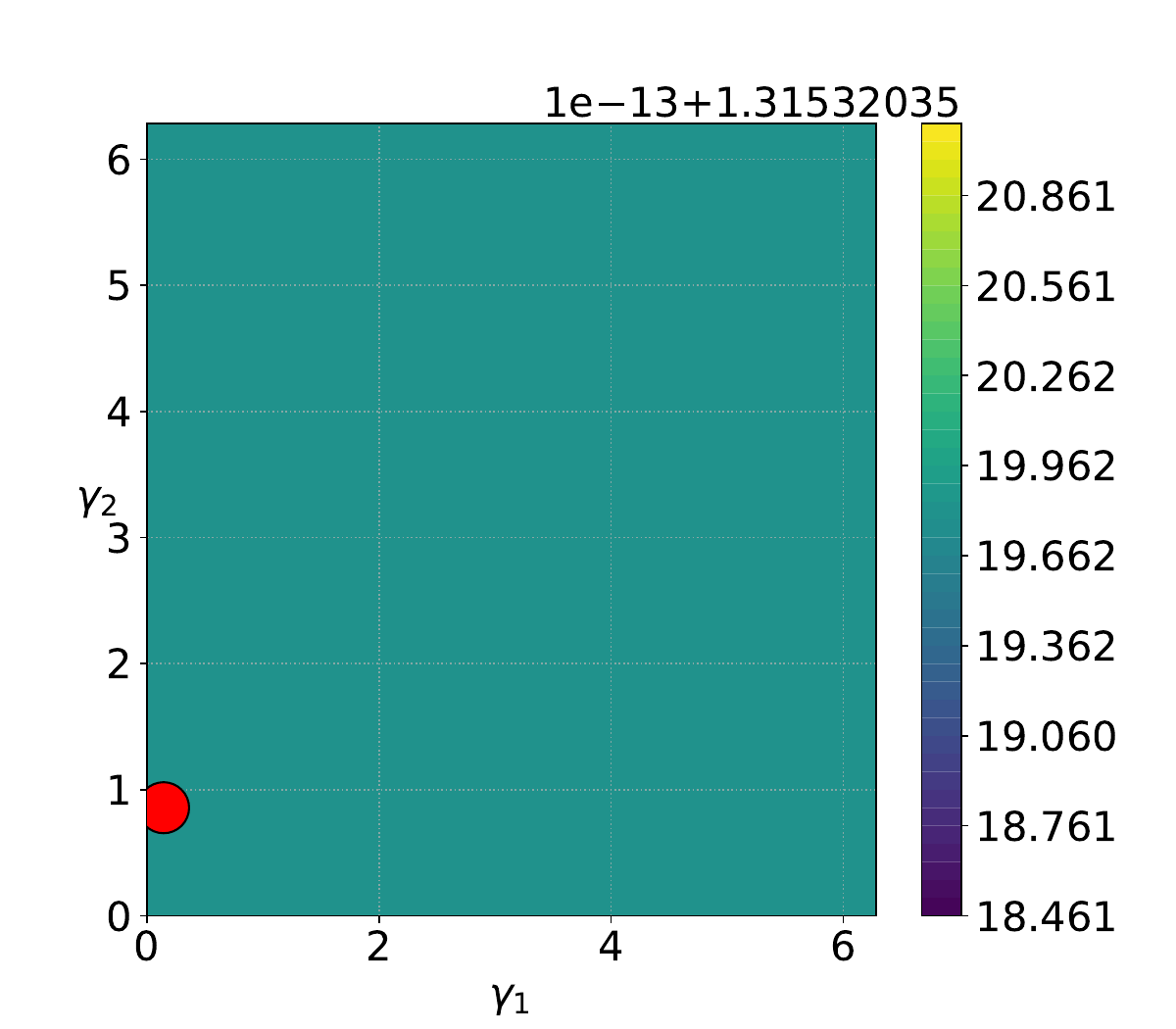}
        \caption{Landscape for $\gamma_1$, $\gamma_2$}
        \label{fig:sv-f}
    \end{subfigure}
    \vspace{-0.5em}
    \caption{Full \ac{qaoa} optimization landscapes (noiseless State Vector simulation) showing pairwise parameter dependencies. Other parameters are fixed at the global minimum (red point).}
    \label{fig:landscape_state_vec}
\end{figure}
Fig. \ref{fig:landscape_state_vec} displays the full \ac{qaoa} optimization landscape under noiseless simulation. Crucially, the objective function cost exhibits near-zero variance along the $\gamma_1$ and $\gamma_2$ axes (subfigures~\ref{fig:sv-b}-\ref{fig:sv-e}). This confirms that, in the noiseless regime, the parameters $\gamma_1$ and $\gamma_2$ do not significantly affect the landscape features or its global minimum, which is the primary motivation for our subsequent parameter-filtered analysis focused on $\beta_1$ and $\beta_2$.

\begin{figure}[htb]
    \centering
    
    \begin{subfigure}[b]{0.28\textwidth}
        \centering
        \includegraphics[width=\textwidth]{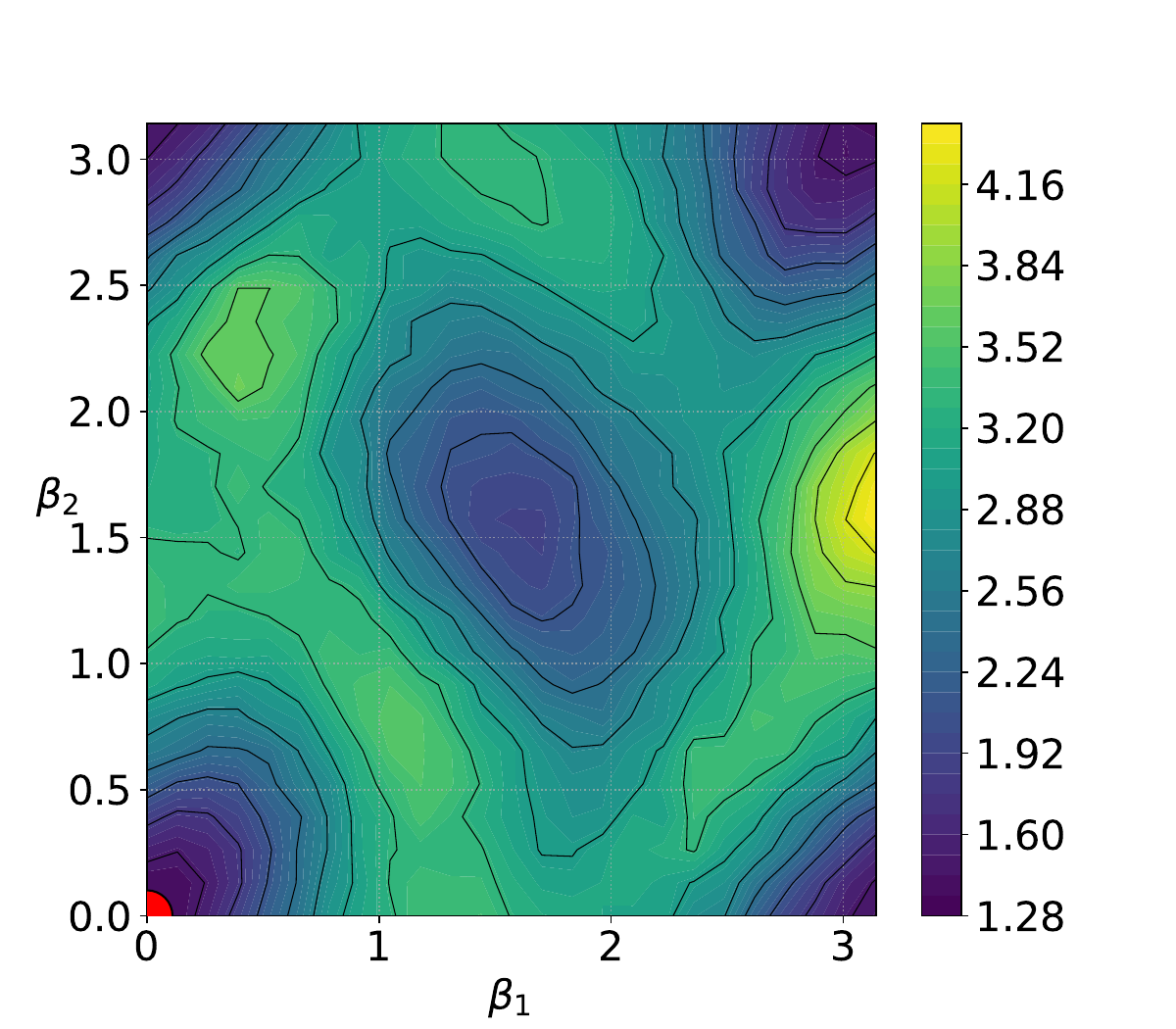}
        \caption{Landscape for $\beta_1$, $\beta_2$}
        \label{fig:sam-a}
    \end{subfigure}
    \hfill 
    \begin{subfigure}[b]{0.28\textwidth}
        \centering
        \includegraphics[width=\textwidth]{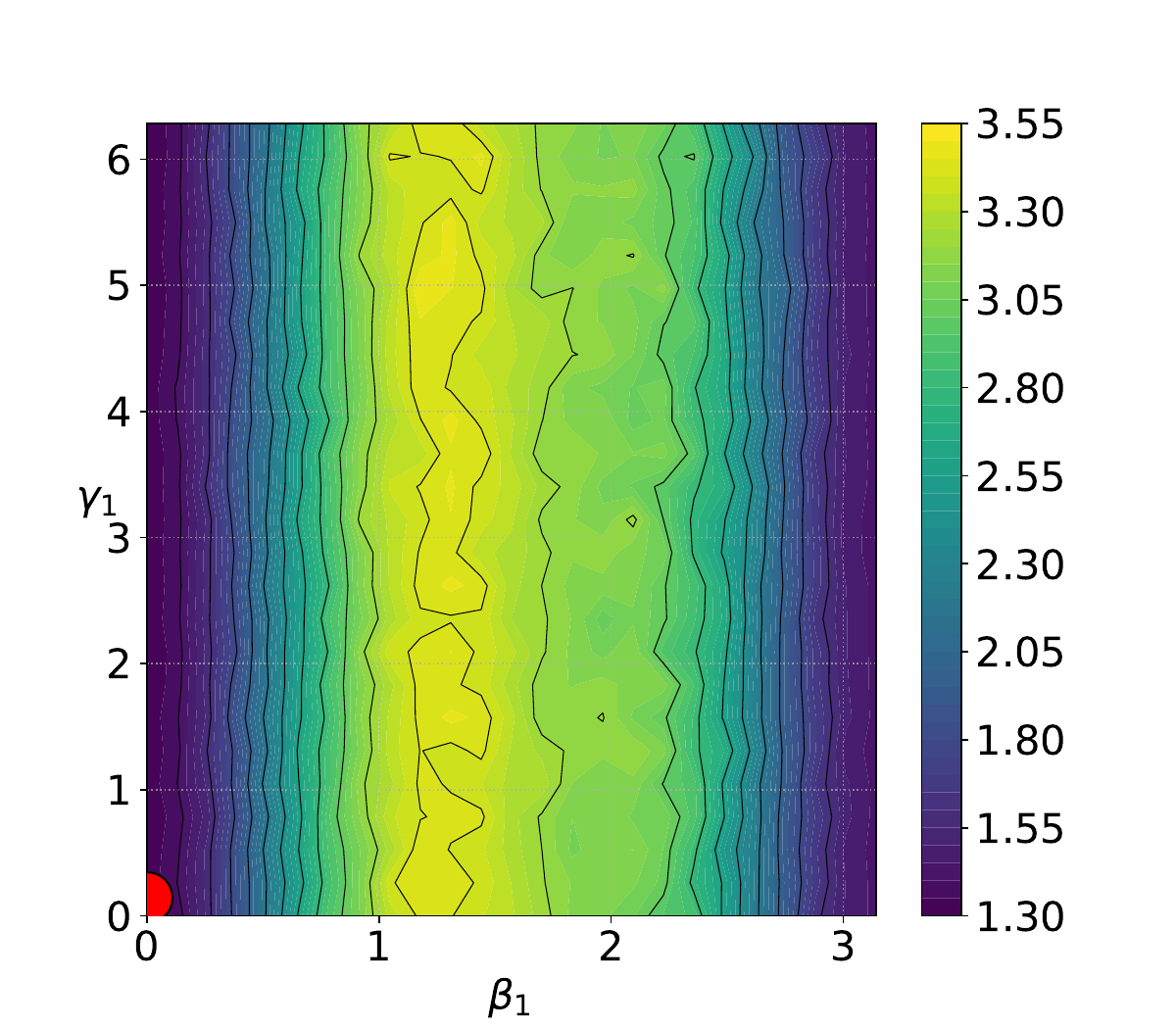}
        \caption{Landscape for $\beta_1$, $\gamma_1$}
        \label{fig:sam-b}
    \end{subfigure}
    \hfill
    \begin{subfigure}[b]{0.28\textwidth}
        \centering
        \includegraphics[width=\textwidth]{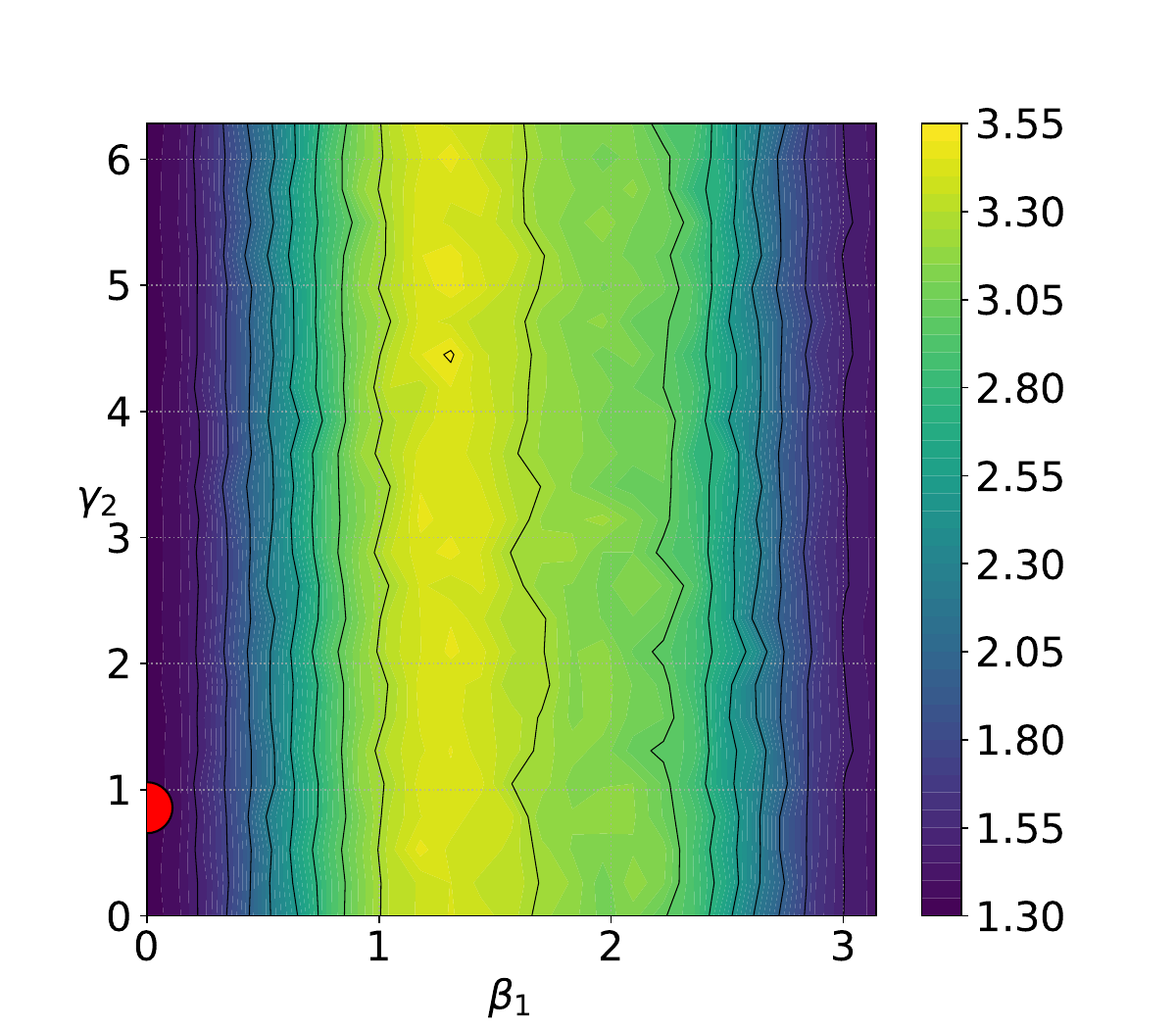}
        \caption{Landscape for $\beta_1$, $\gamma_2$}
        \label{fig:sam-c}
    \end{subfigure}
    
    \begin{subfigure}[b]{0.28\textwidth}
        \centering
        \includegraphics[width=\textwidth]{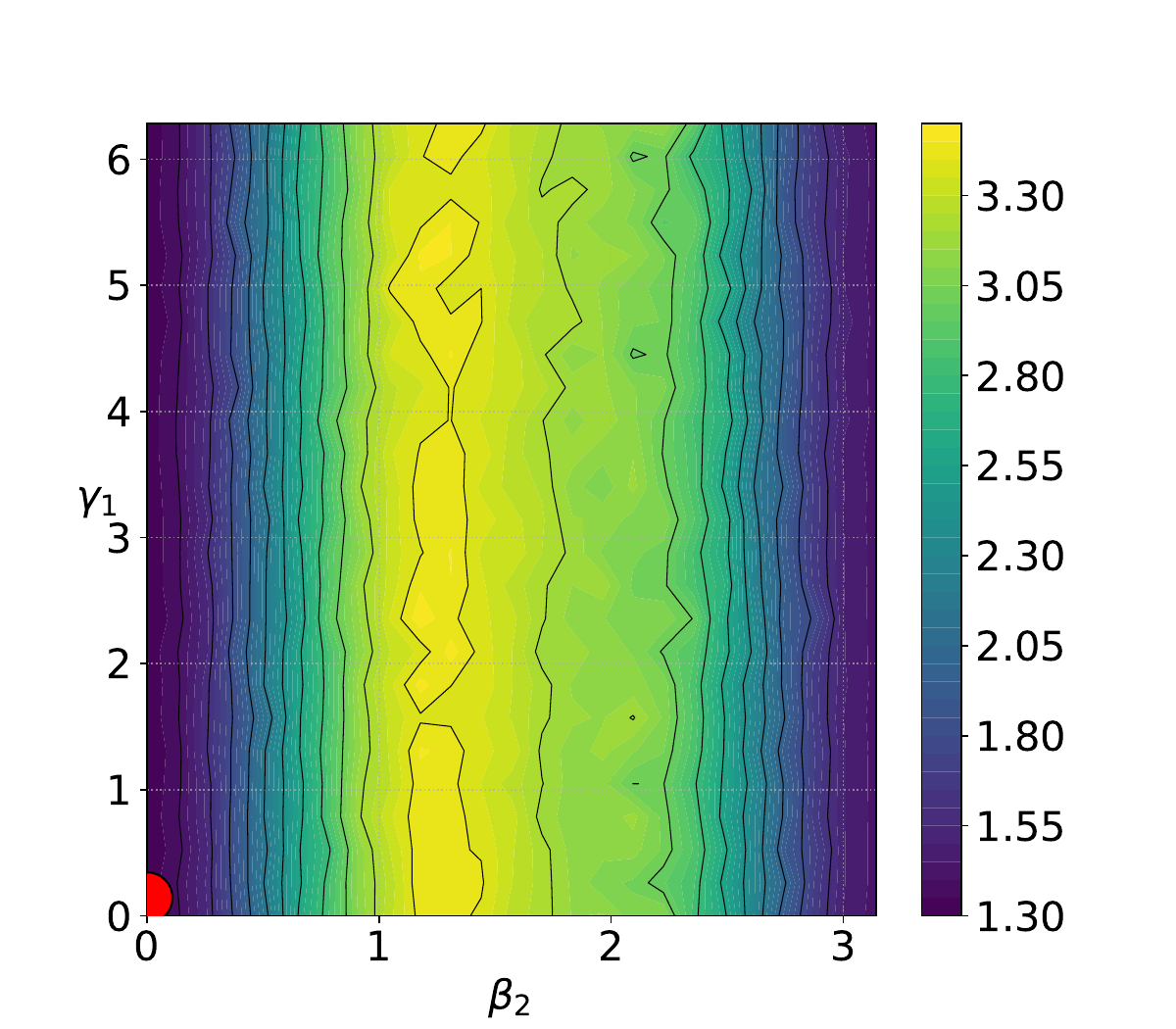}
        \caption{Landscape for $\beta_2$, $\gamma_1$}
        \label{fig:sam-d}
    \end{subfigure}
    \hfill 
    \begin{subfigure}[b]{0.28\textwidth}
        \centering
        \includegraphics[width=\textwidth]{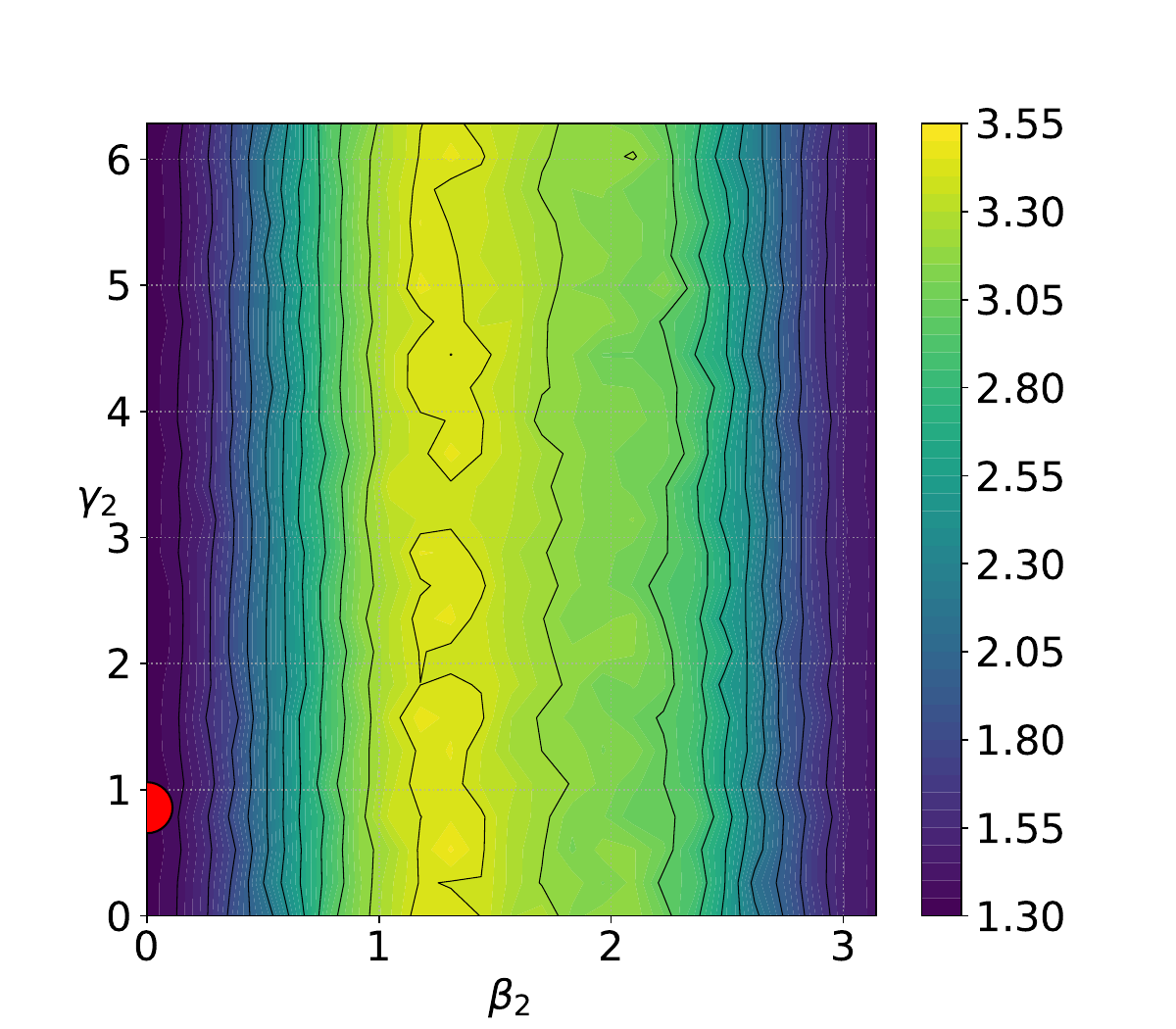}
        \caption{Landscape for $\beta_2$, $\gamma_2$}
        \label{fig:sam-e}
    \end{subfigure}
    \hfill
    \begin{subfigure}[b]{0.28\textwidth}
        \centering
        \includegraphics[width=\textwidth]{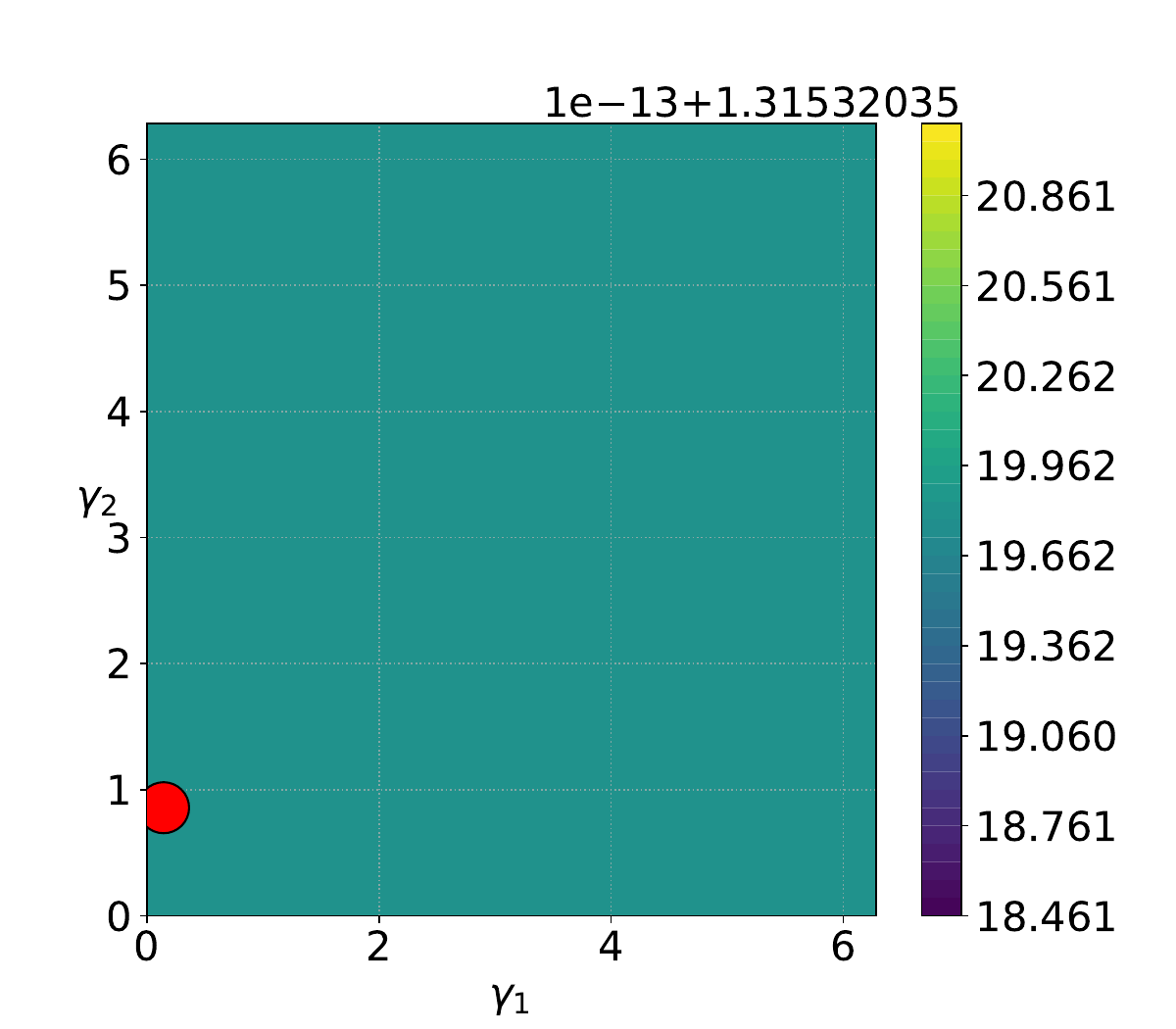}
        \caption{Landscape for $\gamma_1$, $\gamma_2$}
        \label{fig:sam-f}
    \end{subfigure}
    \vspace{-0.5em}
    \caption{Full \ac{qaoa} optimization landscapes (Sampling Noise simulation, 1024 shots) showing pairwise parameter dependencies. Other parameters are fixed at the global minimum (red point).}
    \label{fig:landscape_sampling}
\end{figure}
The introduction of Sampling Noise (Fig. \ref{fig:landscape_sampling}) introduces some localized noise, but the $\gamma_i$ parameters remain largely inactive.

\begin{figure}[htb]
    \centering
    
    \begin{subfigure}[b]{0.28\textwidth}
        \centering
        \includegraphics[width=\textwidth]{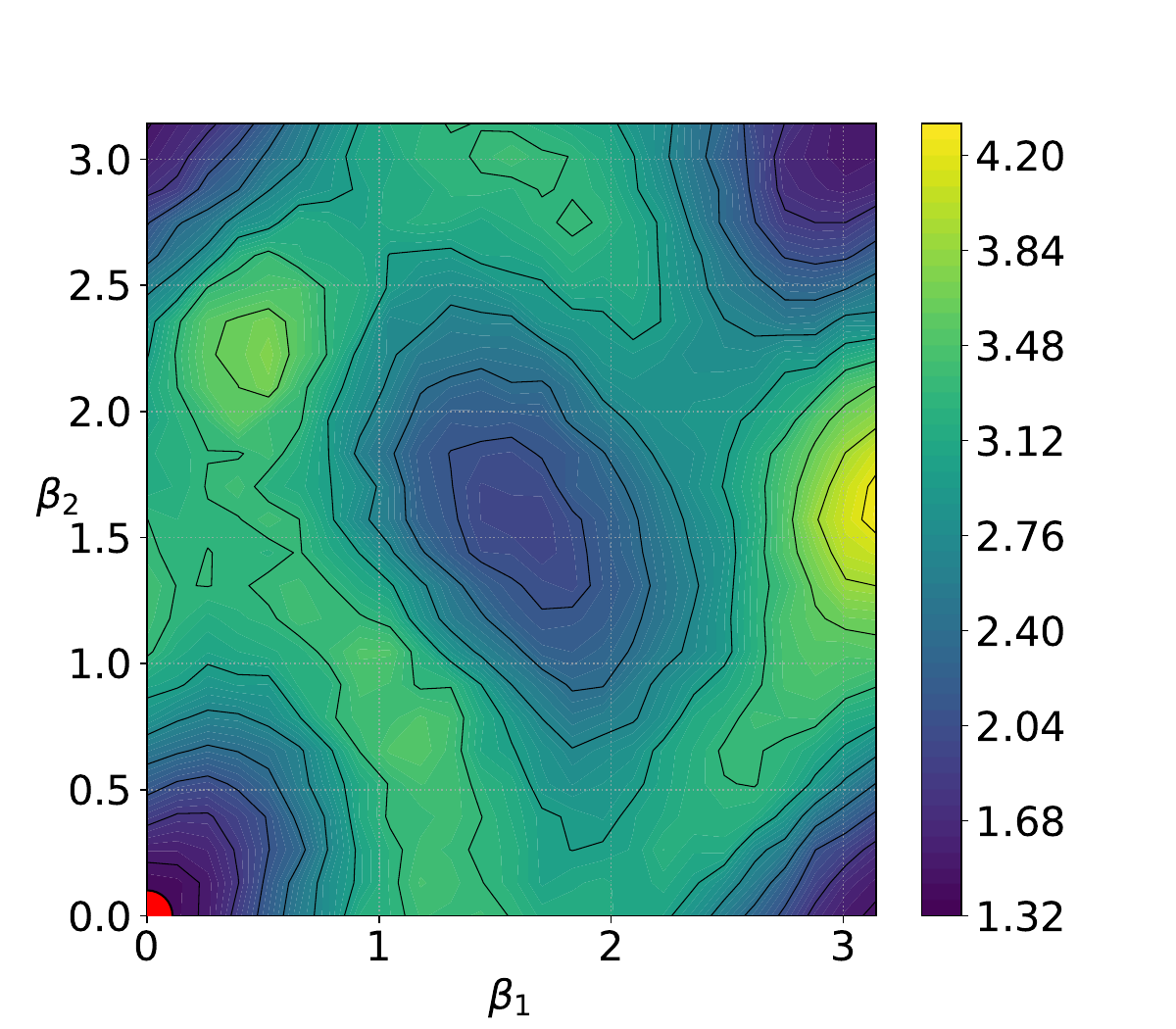}
        \caption{Landscape for $\beta_1$, $\beta_2$}
        \label{fig:tta-a}
    \end{subfigure}
    \hfill 
    \begin{subfigure}[b]{0.28\textwidth}
        \centering
        \includegraphics[width=\textwidth]{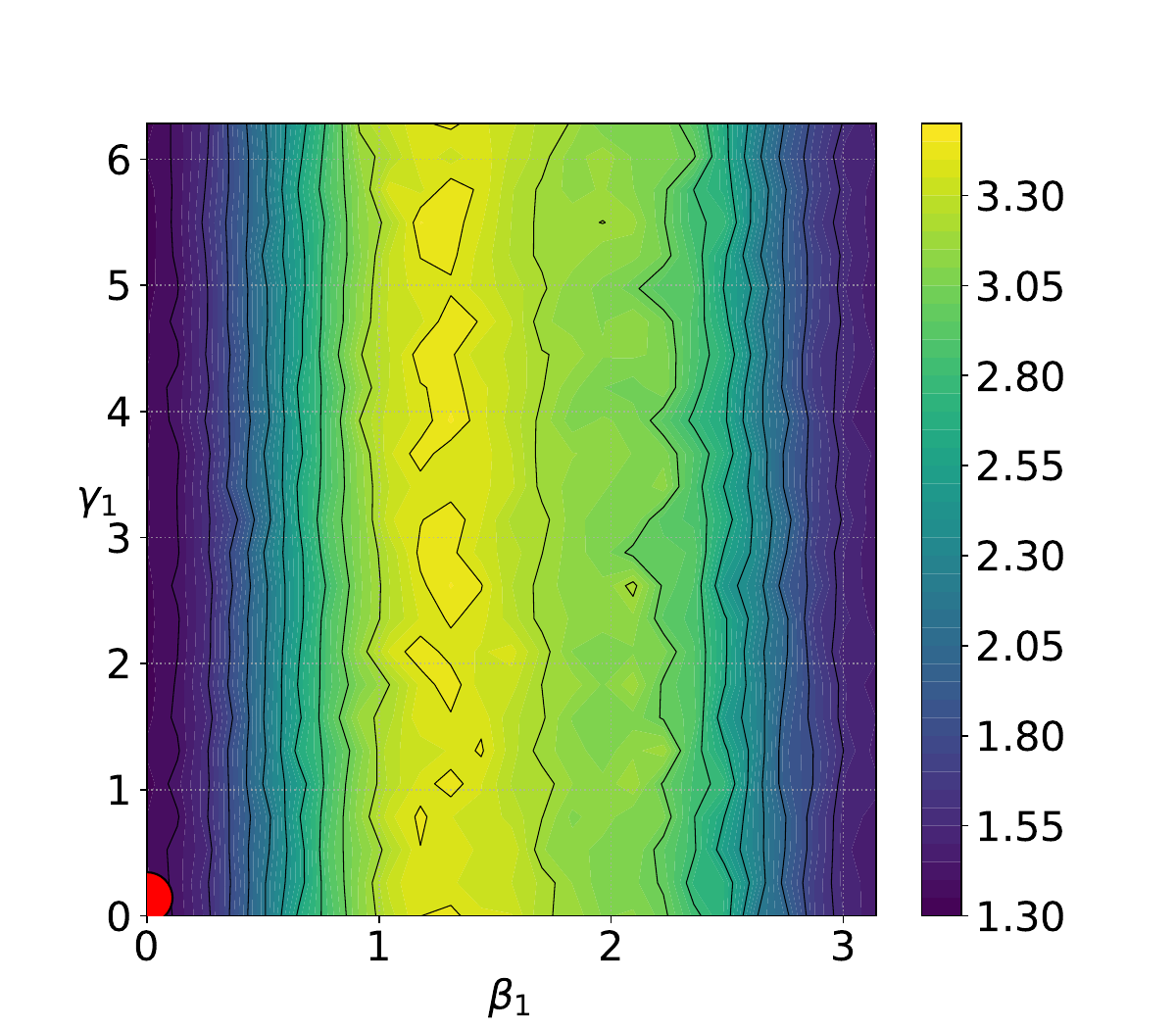}
        \caption{Landscape for $\beta_1$, $\gamma_1$}
        \label{fig:tta-b}
    \end{subfigure}
    \hfill
    \begin{subfigure}[b]{0.28\textwidth}
        \centering
        \includegraphics[width=\textwidth]{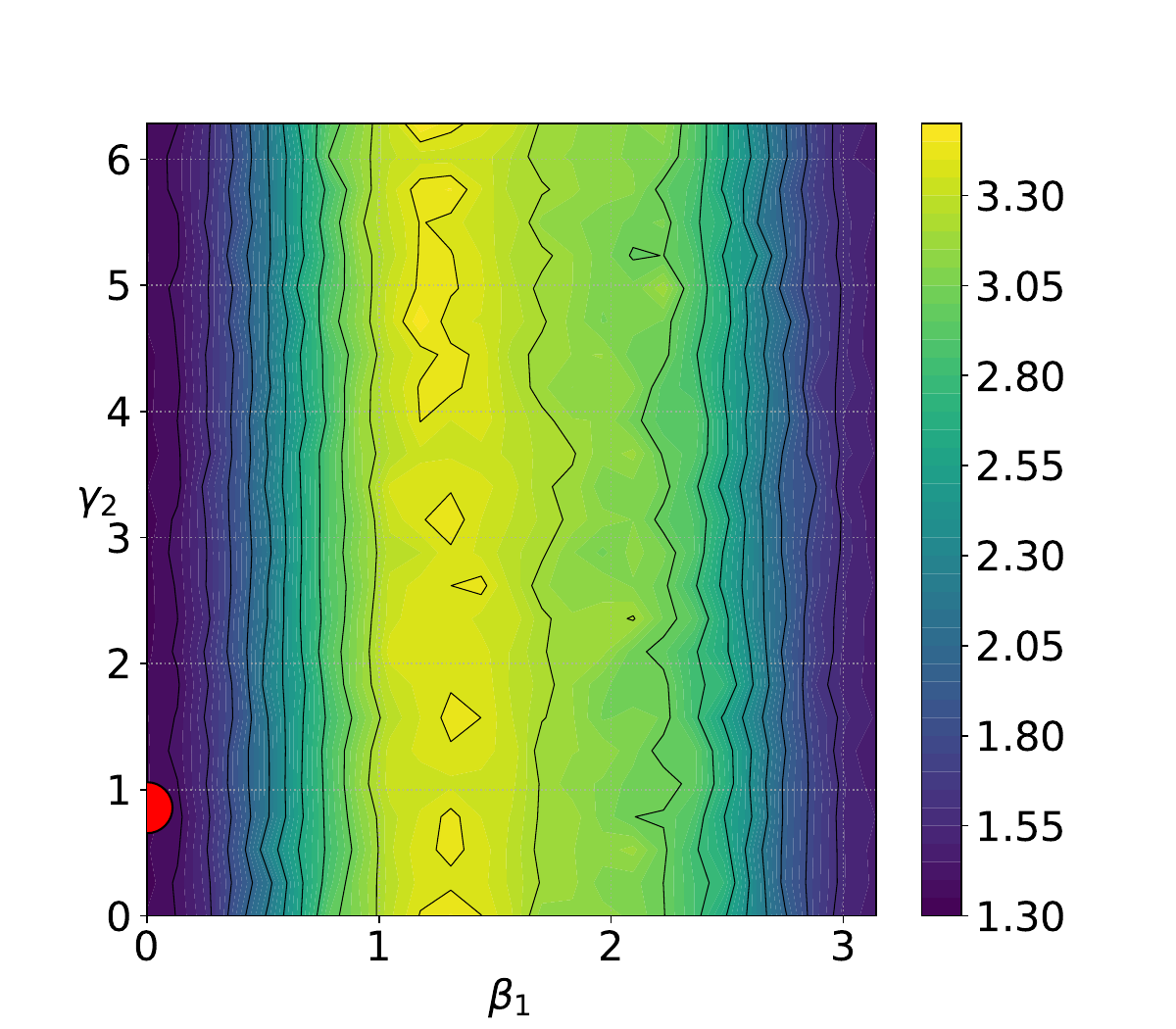}
        \caption{Landscape for $\beta_1$, $\gamma_2$}
        \label{fig:tta-c}
    \end{subfigure}
    
    \begin{subfigure}[b]{0.28\textwidth}
        \centering
        \includegraphics[width=\textwidth]{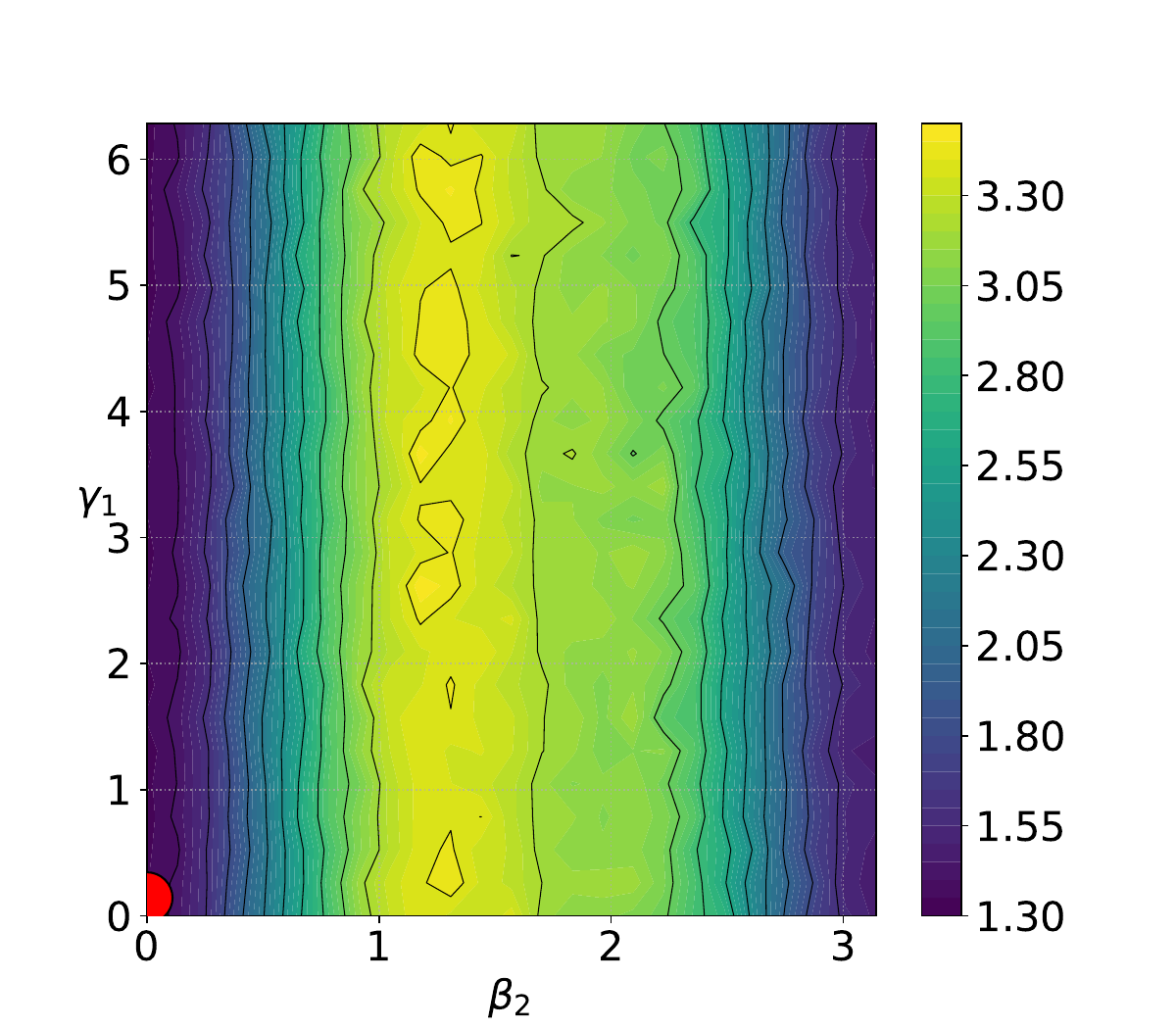}
        \caption{Landscape for $\beta_2$, $\gamma_1$}
        \label{fig:tta-d}
    \end{subfigure}
    \hfill 
    \begin{subfigure}[b]{0.28\textwidth}
        \centering
        \includegraphics[width=\textwidth]{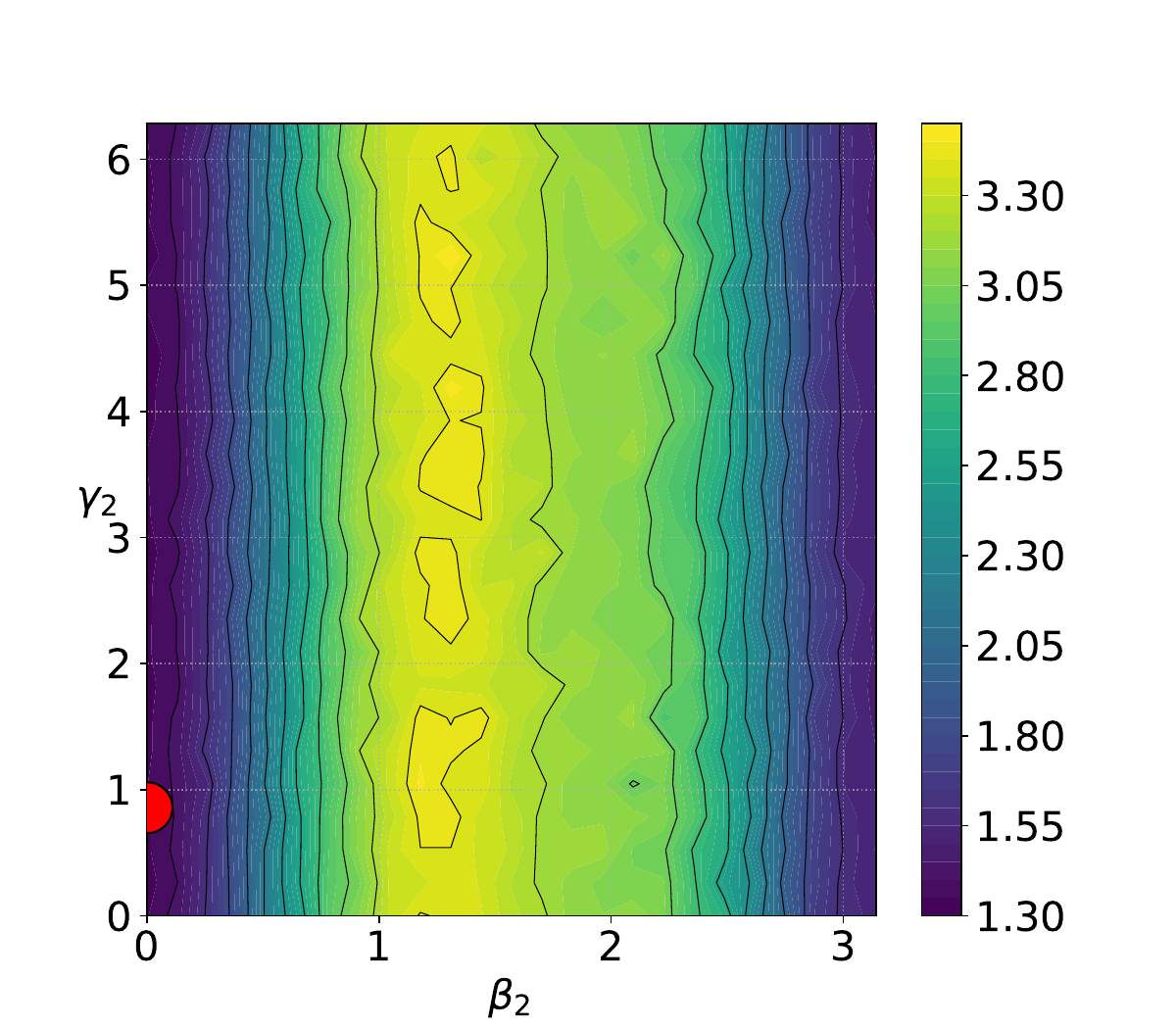}
        \caption{Landscape for $\beta_2$, $\gamma_2$}
        \label{fig:tta-e}
    \end{subfigure}
    \hfill
    \begin{subfigure}[b]{0.28\textwidth}
        \centering
        \includegraphics[width=\textwidth]{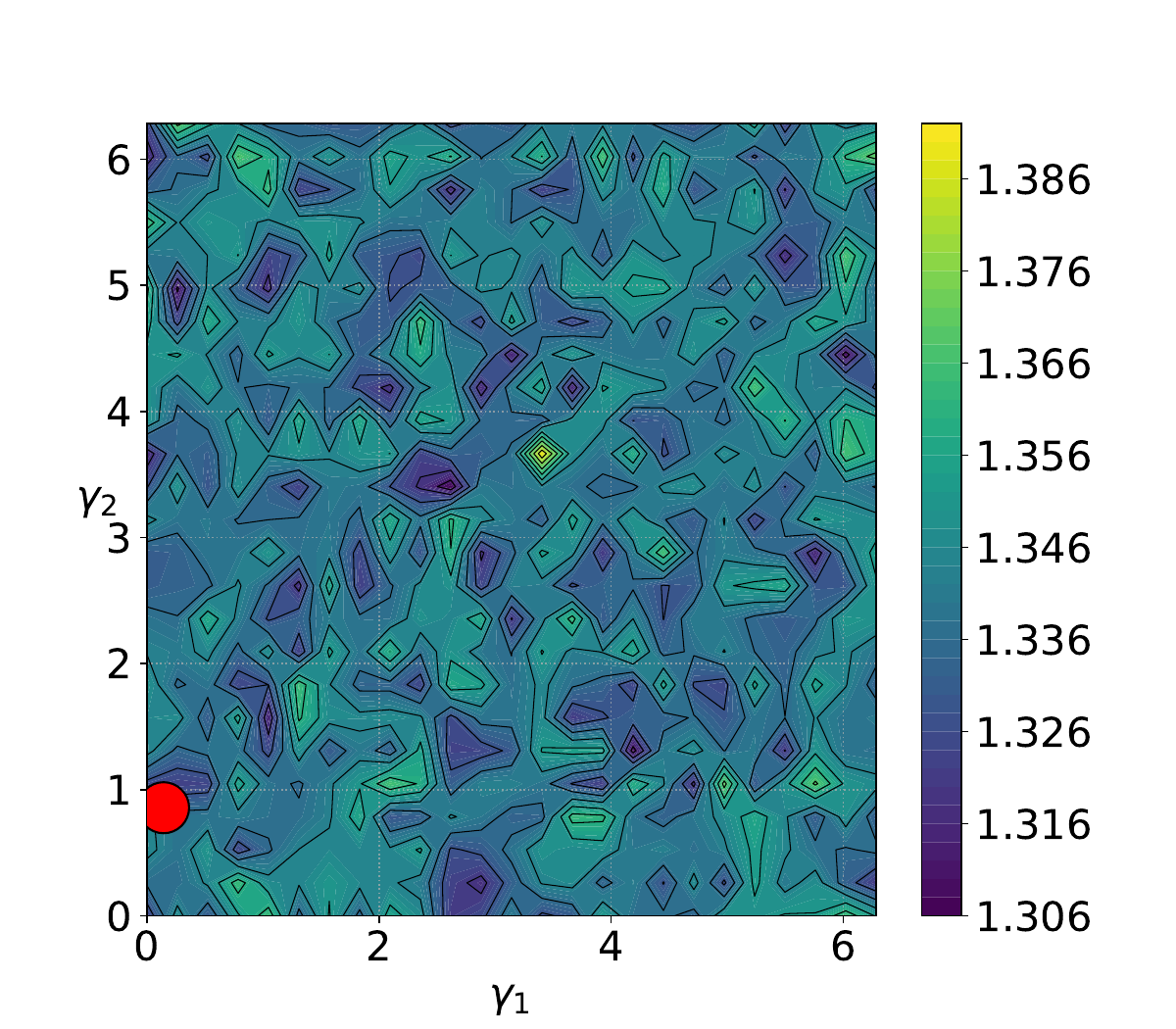}
        \caption{Landscape for $\gamma_1$, $\gamma_2$}
        \label{fig:tta-f}
    \end{subfigure}
    \vspace{-0.5em}
    \caption{Full \ac{qaoa} optimization landscapes (Thermal Noise-A: $T_1 = 380\mu s, T_2 = 400 \mu s$, 1024 shots) showing pairwise parameter dependencies. Other parameters are fixed at the global minimum (red point).}
    \label{fig:landscape_tta}
\end{figure}

\begin{figure}[htb]
    \centering
    
    \begin{subfigure}[b]{0.28\textwidth}
        \centering
        \includegraphics[width=\textwidth]{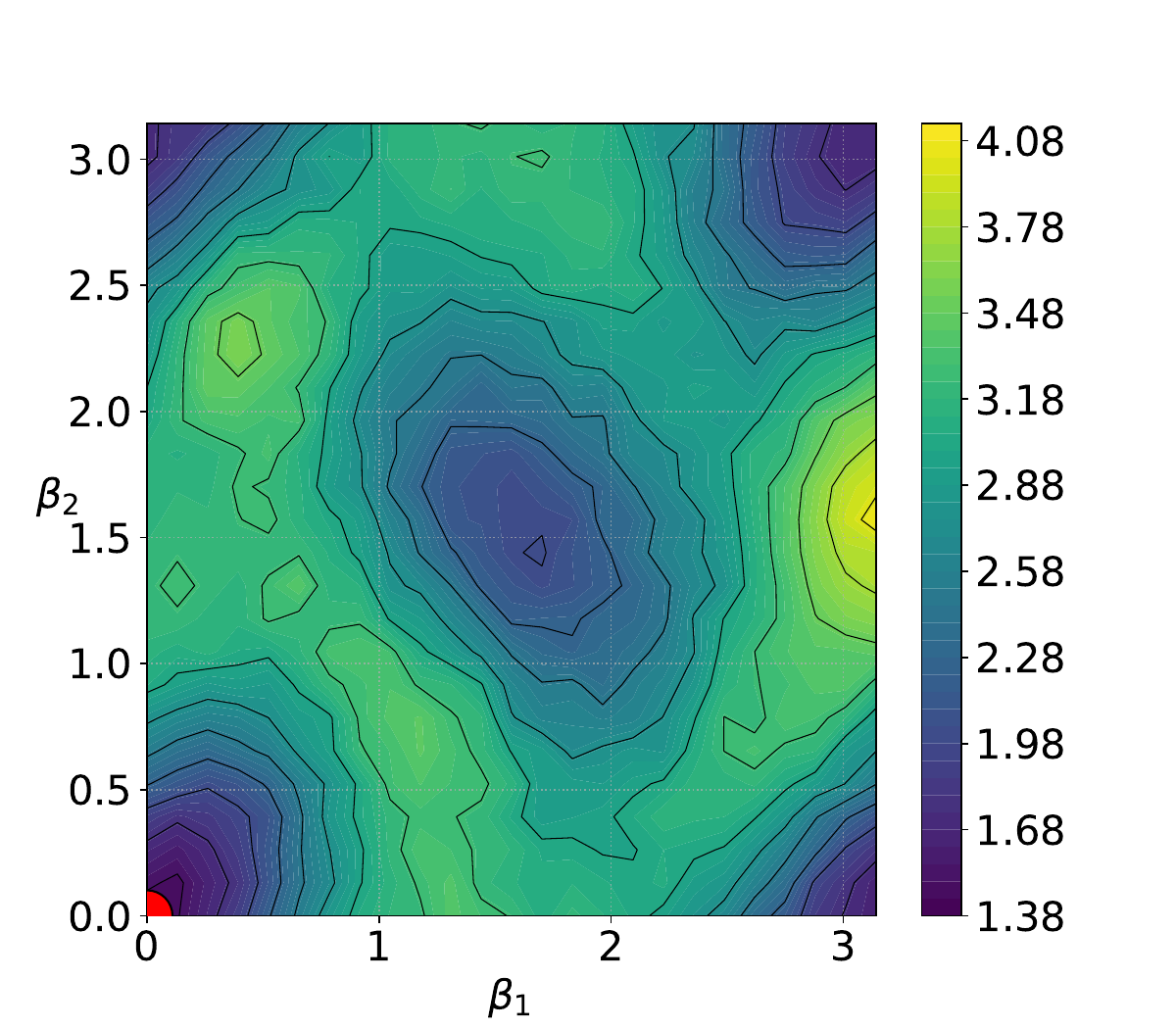}
        \caption{Landscape for $\beta_1$, $\beta_2$}
        \label{fig:ttb-a}
    \end{subfigure}
    \hfill 
    \begin{subfigure}[b]{0.28\textwidth}
        \centering
        \includegraphics[width=\textwidth]{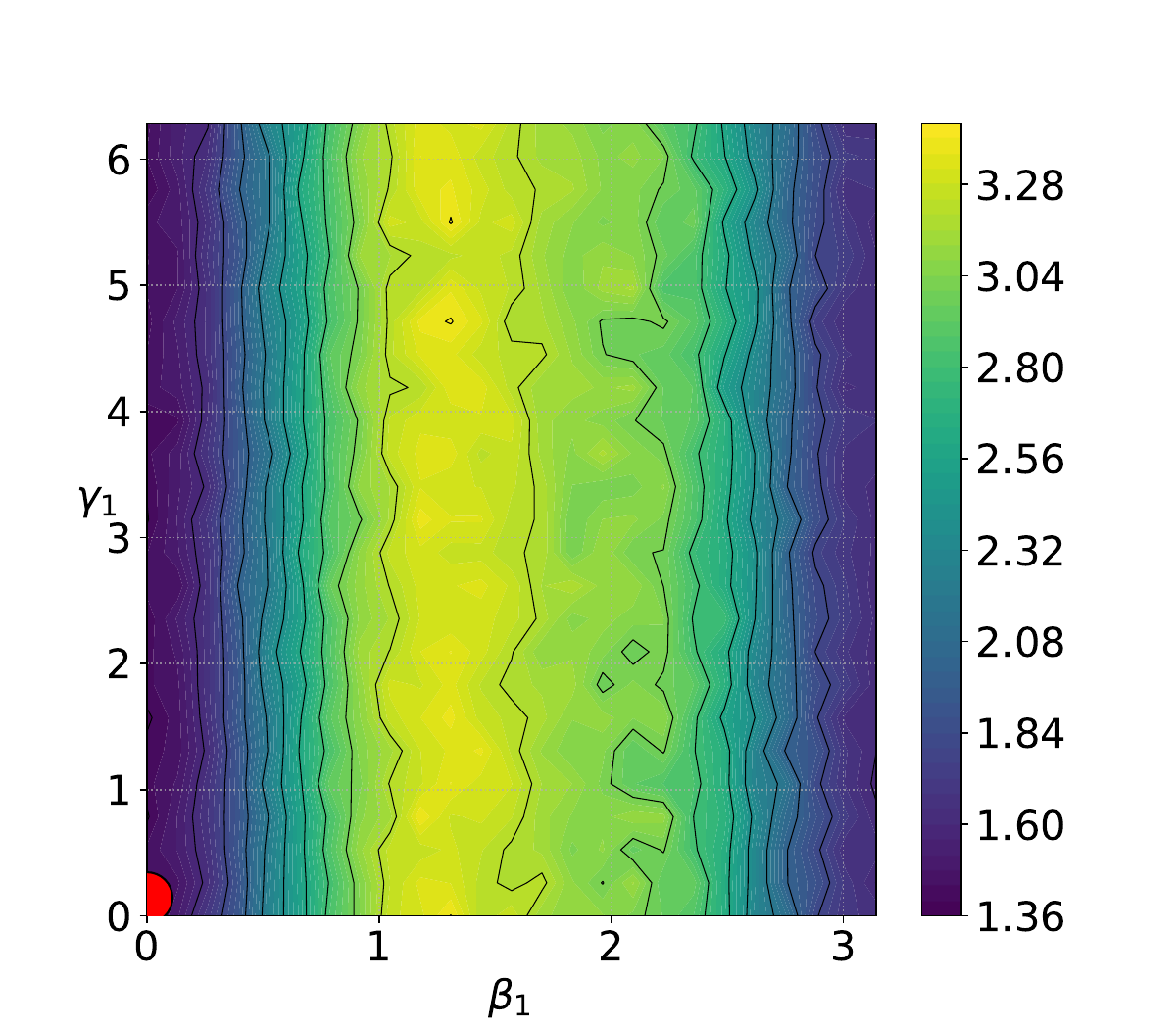}
        \caption{Landscape for $\beta_1$, $\gamma_1$}
        \label{fig:ttb-b}
    \end{subfigure}
    \hfill
    \begin{subfigure}[b]{0.28\textwidth}
        \centering
        \includegraphics[width=\textwidth]{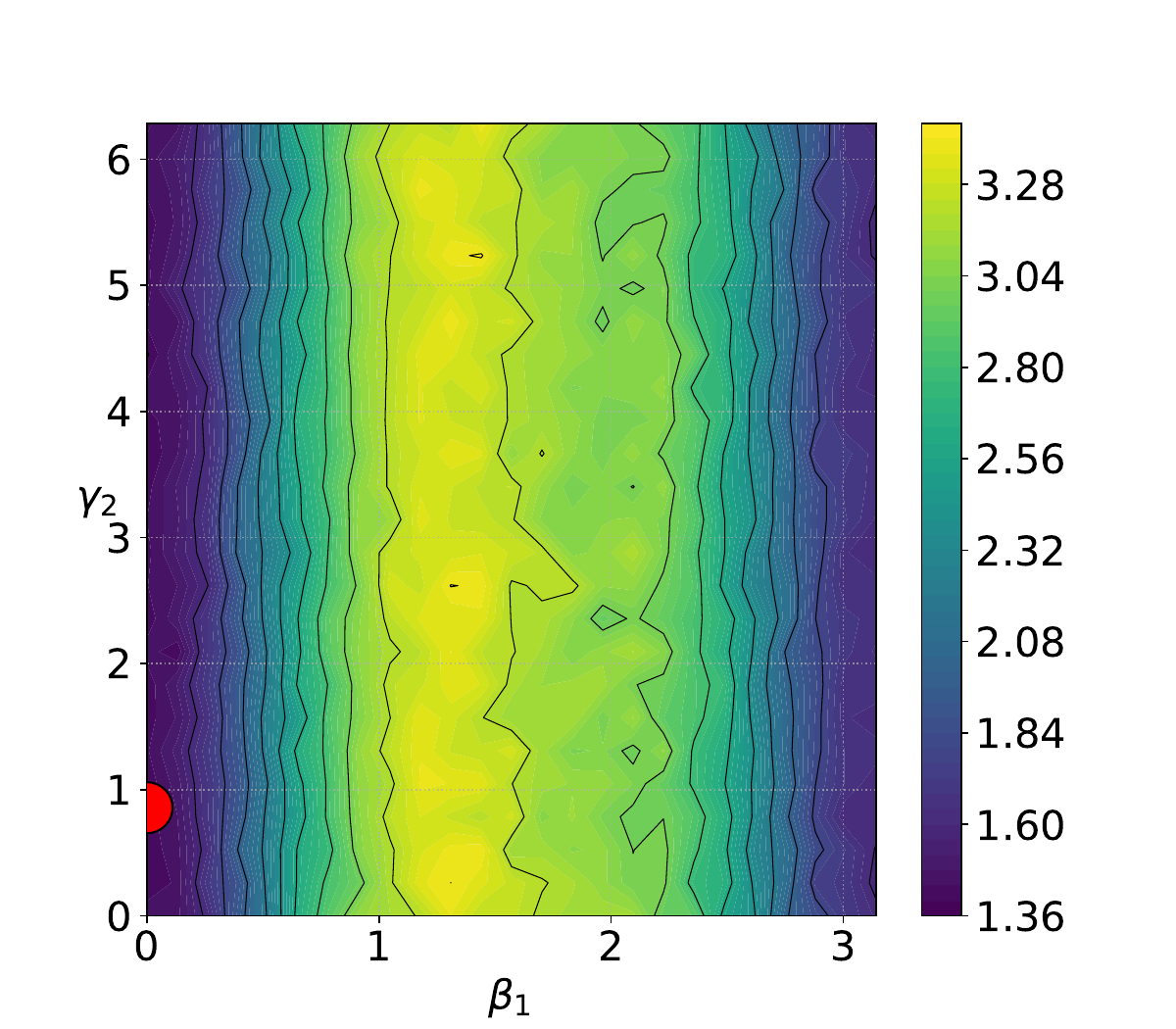}
        \caption{Landscape for $\beta_1$, $\gamma_2$}
        \label{fig:ttb-c}
    \end{subfigure}
    
    \begin{subfigure}[b]{0.28\textwidth}
        \centering
        \includegraphics[width=\textwidth]{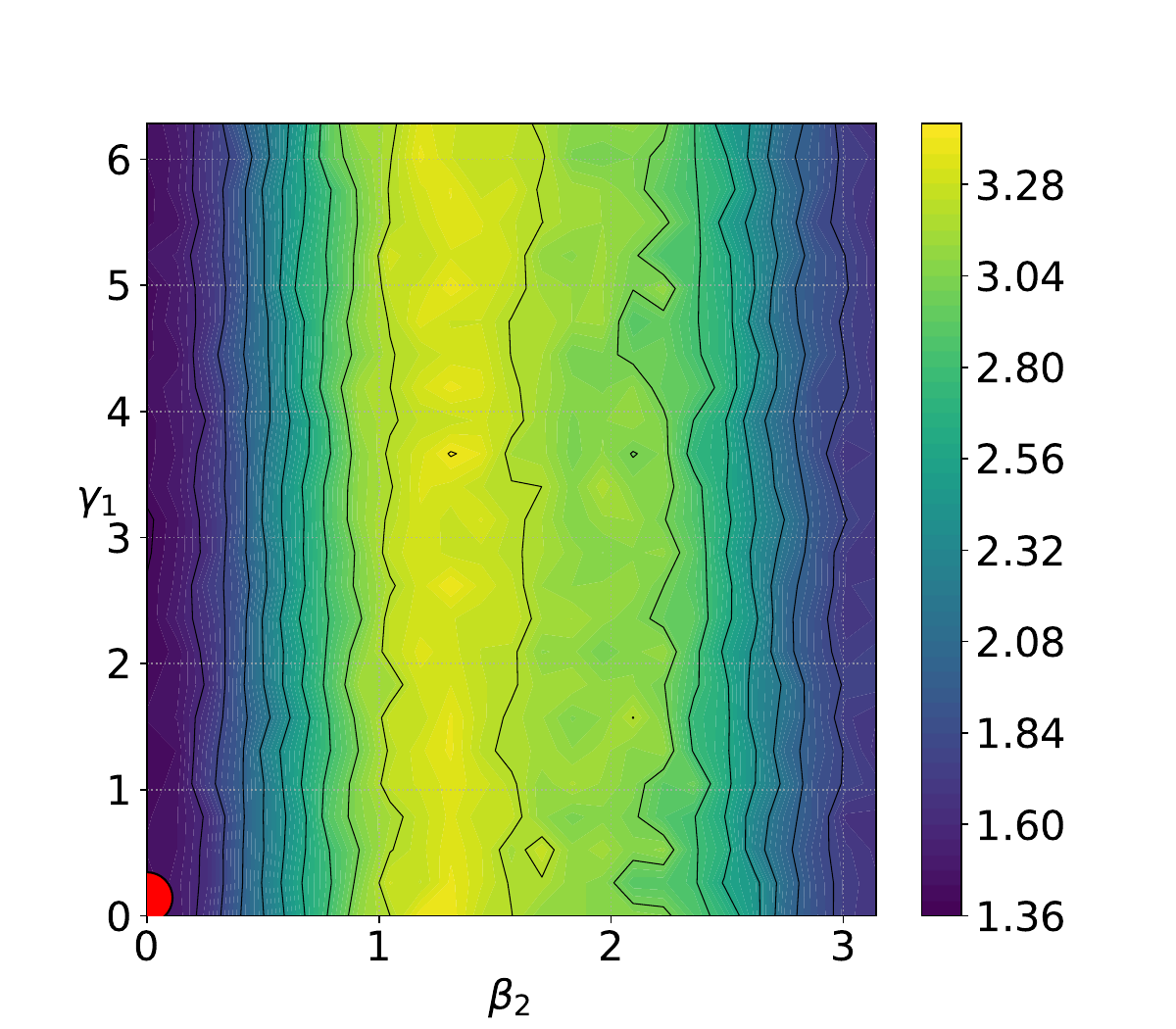}
        \caption{Landscape for $\beta_2$, $\gamma_1$}
        \label{fig:ttb-d}
    \end{subfigure}
    \hfill 
    \begin{subfigure}[b]{0.28\textwidth}
        \centering
        \includegraphics[width=\textwidth]{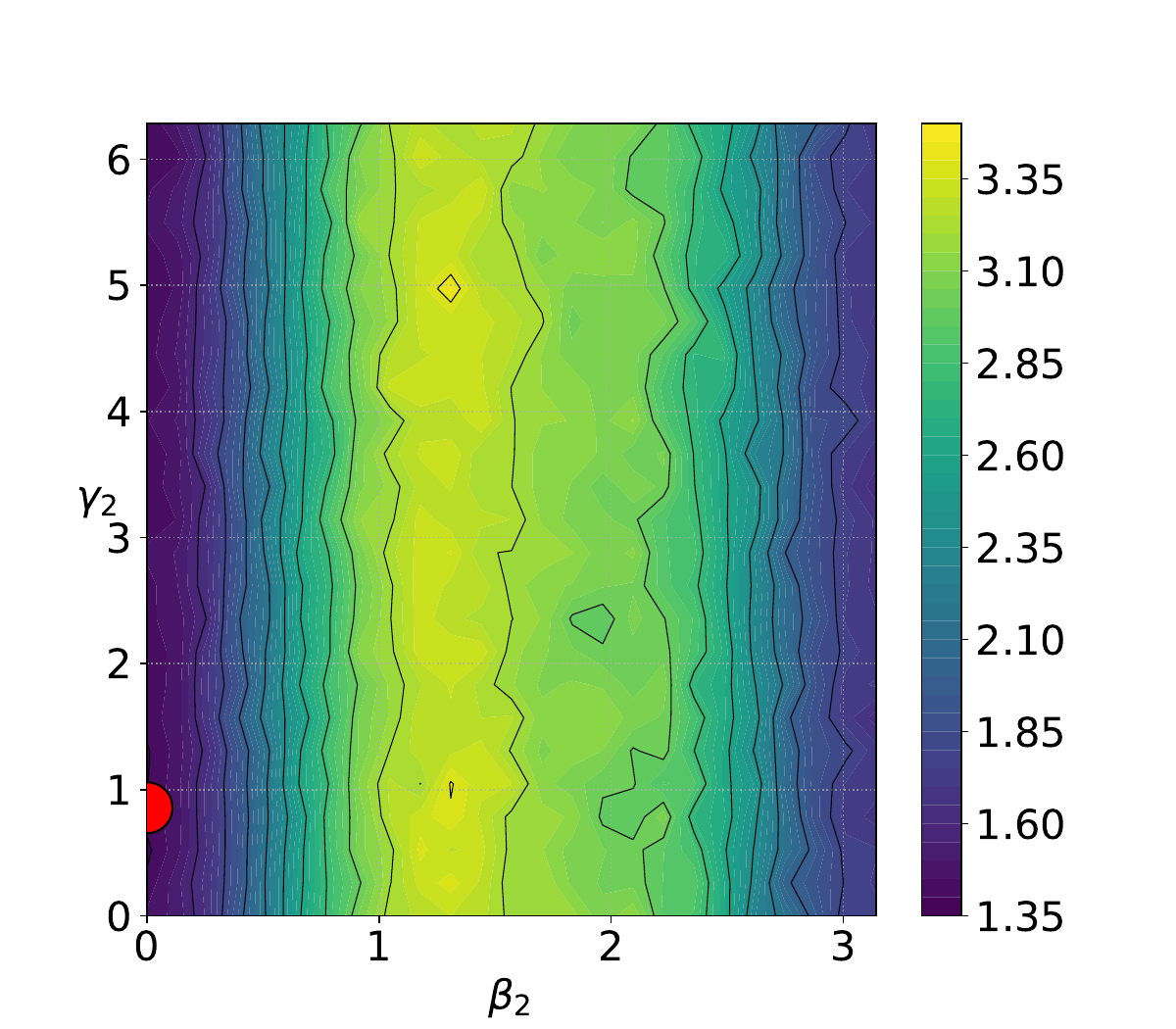}
        \caption{Landscape for $\beta_2$, $\gamma_2$}
        \label{fig:ttb-e}
    \end{subfigure}
    \hfill
    \begin{subfigure}[b]{0.28\textwidth}
        \centering
        \includegraphics[width=\textwidth]{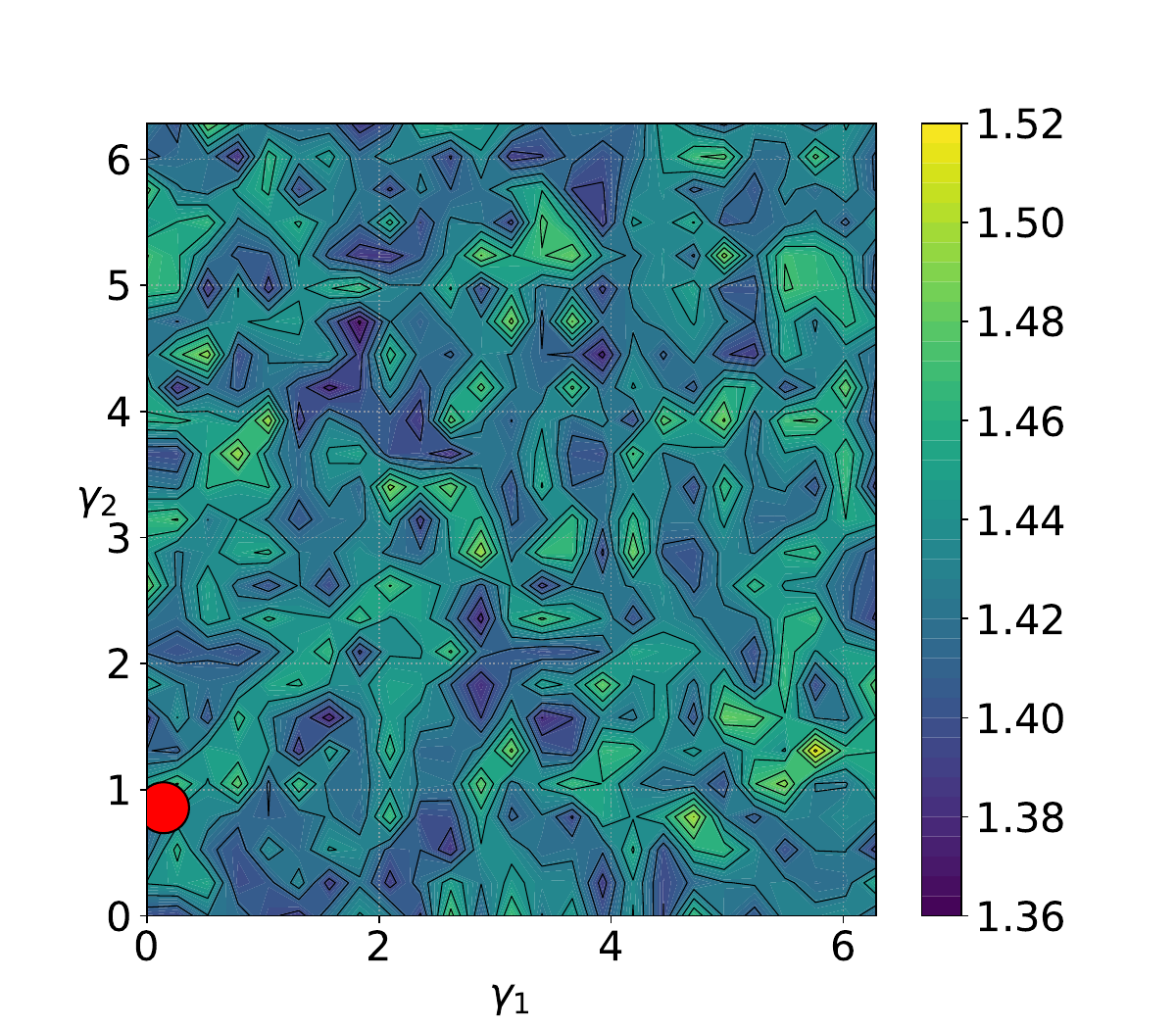}
        \caption{Landscape for $\gamma_1$, $\gamma_2$}
        \label{fig:ttb-f}
    \end{subfigure}
    \vspace{-0.5em}
    \caption{Full \ac{qaoa} optimization landscapes (Thermal Noise-A: $T_1 = 80\mu s, T_2 = 100 \mu s$, 1024 shots) showing pairwise parameter dependencies. Other parameters are fixed at the global minimum (red point).}
    \label{fig:landscape_ttb}
\end{figure}
With the introduction of Thermal Noise-A and particularly Thermal Noise-B, the $\beta_1$ and $\beta_2$ landscape (subfigures~\ref{fig:tta-a} and~\ref{fig:ttb-a}) becomes increasingly rugged and multi-modal. Furthermore, unlike the noiseless case, the $\gamma_1, \gamma_2$ landscape (subfigures~\ref{fig:tta-f} and ~\ref{fig:ttb-f}) under thermal noise exhibits significant noise and non-convex features, suggesting that noise can activate previously inert parameters or severely corrupt their optimization subspaces.

\subsection{Optimization Results}
The optimizers were assessed based on 10 independent runs per scenario (using random initial parameters) under two approaches, driven by the landscape analysis (Sec.~\ref{sec:landscape}): standard optimization (optimizing all four parameters $\beta_1, \beta_2, \gamma_1, \gamma_2 $) and parameter-filtered optimization (optimizing only $\beta_1,\beta_2$, with $\gamma_1,\gamma_2$ fixed to their initial values). 

\begin{figure}[htbp]
    \centering
    
    \begin{subfigure}[b]{\textwidth}
        \centering
        \includegraphics[width=\textwidth]{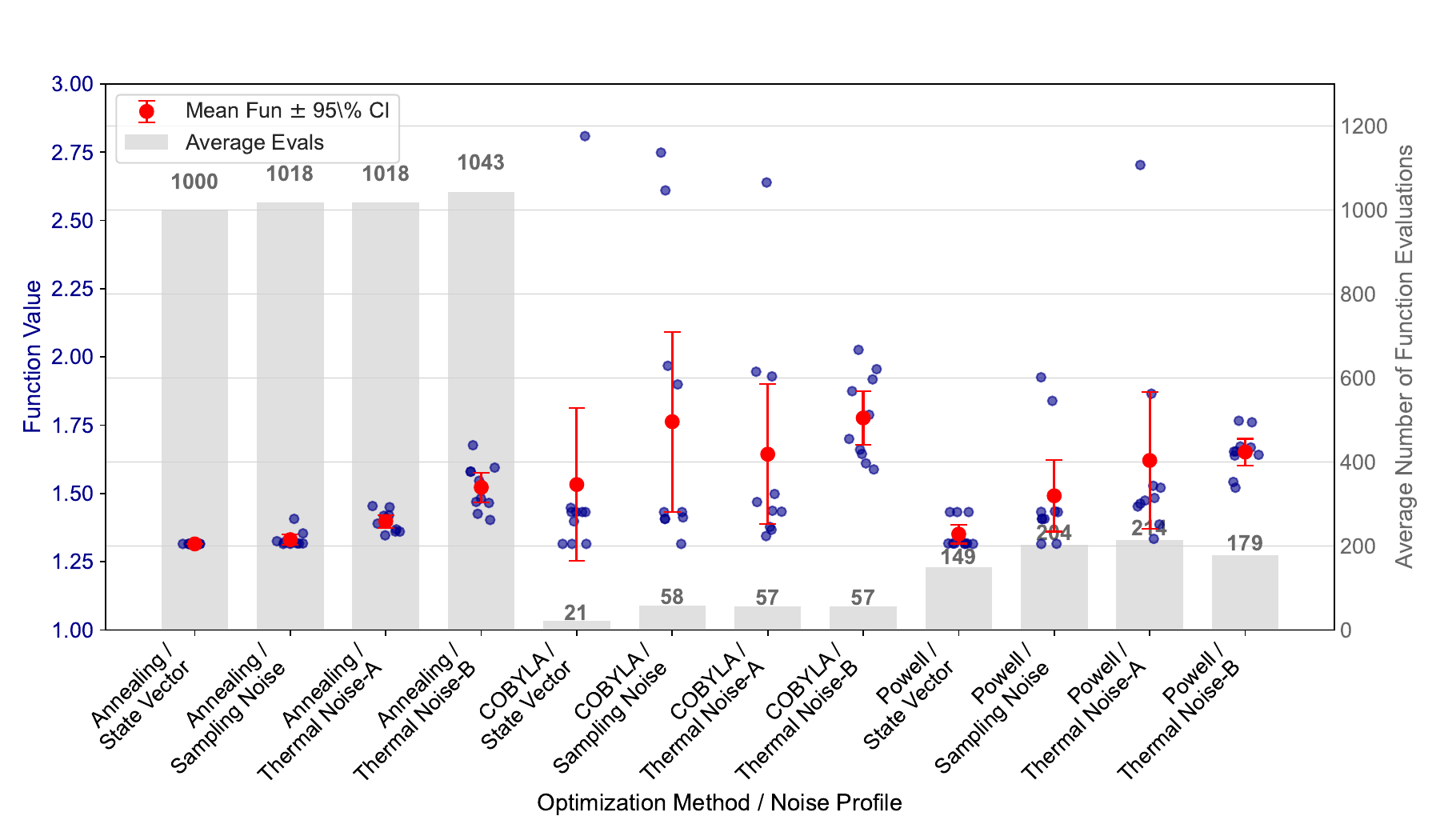}
        \caption{Standard Optimization: All four parameters ($\beta_1$, $\beta_2, \gamma_1, \gamma_2$) were optimized. }
        \label{fig:optimizers_both}
    \end{subfigure}
    \begin{subfigure}[b]{\textwidth}
        \centering
        \includegraphics[width=\textwidth]{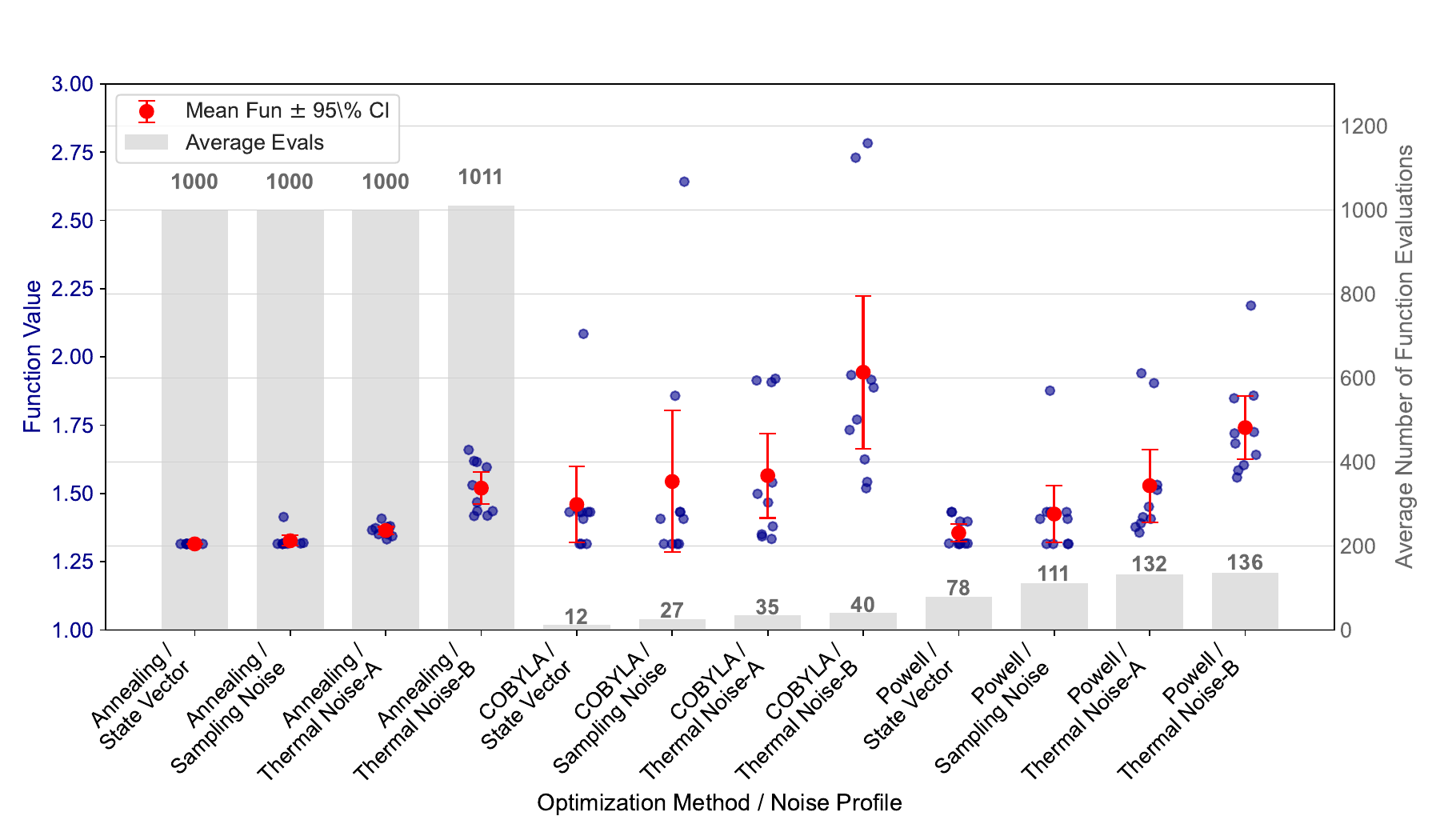}
        \caption{Parameter-Filtered Optimization: Optimization restricted to active parameters ($\beta_1$, $\beta_2$).}
        \label{fig:optimizers_beta}
    \end{subfigure}
    \caption{
    Combined \ac{qaoa} Convergence Benchmark of Classical Optimizers under four noise profiles (Sec.~\ref{sec:noise_profiles}). (a) shows results from the standard optimization approach, and (b) shows the results from the parameter-filtered approach. Blue dots show 10 individual runs (Primary Y-axis: Function Value). Red markers and error bars show the mean and $95\%$ confidence interval. Gray bars (Secondary Y-axis) indicate the average number of function evaluations.
    }
    \label{fig:optimizers}
\end{figure}
Figure~\ref{fig:optimizers_both} shows the standard optimization results. The exhaustive Dual Annealing required the maximum function evaluations and achieved the lowest noiseless cost ($\approx1.26$), and demonstrated the greatest stability under noise, maintaining a tighter performance distribution and better final results than the other optimizers. Conversely, the highly efficient \ac{cobyla} and Powell Method traded this superior stability and solution quality for significantly fewer function evaluations across all noisy regimes.

Figure~\ref{fig:optimizers_beta} displays the results when optimizers were restricted to the active subspace, optimizing only $\beta_1$ and $\beta_2$, as motivated by the landscape analysis (Sec.~\ref{sec:landscape}).
The most notable outcome was the substantial efficiency gain for the fast optimizers: \ac{cobyla} achieved its best results with only 12 function evaluations, Powell Method with only 78 (a significant reduction from the four-parameter search). While Dual Annealing's efficiency remained unchanged, restricting the search space resulted in improved robustness for the fast optimizers. Specifically, \ac{cobyla}'s and Powell Method's mean function value and variance under Sampling Noise and Thermal Noise-A were lower and tighter than in the four-parameter case, demonstrating that filtering inactive parameters is an effective noise mitigation strategy.

\section{Conclusion}
In this work, we conducted a systematic study of classical optimization strategies for \ac{qaoa} applied to \ac{gmvp} under various noise profiles: noiseless State Vector simulation, sampling noise, and two distinct thermal noise models. Our analysis, which benchmarked Dual Annealing, \ac{cobyla}, and the Powell Method, provided critical insights into optimizing \ac{qaoa} performance in the near-term \ac{nisq} era.

\subsection{Key Findings and Parameter Efficiency}
The main discovery arose from the Cost Function Landscape Analysis, which conclusively demonstrated that, in the noiseless regime, the \ac{qaoa} angle parameters $\gamma_1$ and $\gamma_2$ were largely inactive. While the landscape became increasingly rugged and multi-modal under thermal noise, the $\gamma_i$-subspaces remained relatively flat, affirming our methodology. This insight motivated our parameter-filtered approach, which revealed significant performance advantages. By reducing the search space to the active $\beta_1$ and $\beta_2$ parameters, we achieved substantial gains in parameter efficiency (e.g., \ac{cobyla} required 12 evaluations vs. 21 in the standard case). Crucially, this filtering approach also led to improved solution quality and reduced variance under sampling noise and moderated thermal noise for the faster optimizers. In summary, the parameter-filtered approach is essential for the \ac{gmvp} instance studied, enabling fast, robust convergence by leveraging structural insights as an effective architecture-aware noise mitigation technique.

\subsection{Future Work}
While this study provides a concrete solution for the specific \ac{gmvp} instance examined, further research is necessary to generalize these findings. The immediate goal is to analyze additional \ac{qaoa} \ac{gmvp} instances, specifically by varying problem complexity. This generalization will involve investigating how parameter activity and filter efficacy change as the number of \ac{qaoa} layers increases, and analyzing problems with different underlying correlation structures. By broadening this systematic benchmark, we aim to provide universally applicable guidelines for selecting the optimal optimizer and parameter subspace for \ac{vqas} in the \ac{nisq} era.

\clearpage

\section*{Acknowledgments}
Martin Beseda is supported by Italian Government (Ministero dell'Università e della Ricerca, PRIN 2022 PNRR) -- cod.P2022SELA7: ''RECHARGE: monitoRing, tEsting, and CHaracterization of performAnce Regressions`` -- Decreto Direttoriale n. 1205 del 28/7/2023. Vojtěch Novák is supported by Grant of SGS No. SP2025/072, VSB-Technical University of Ostrava, Czech Republic.
The authors thank Chris Long of the Cavendish Laboratory, Department of Physics, University of Cambridge, for helpful discussions regarding \ac{qaoa}.

%


%
%
\bibliographystyle{splncs04}
\bibliography{references,optiq}

\end{document}